%% file: final4bps.tex
\newcommand{\pa}{\partial}
\newcommand{\be}{\begin{equation}}
\newcommand{\ee}{\end{equation}}
\newcommand{\bea}{\begin{eqnarray}}
\newcommand{\eea}{\end{eqnarray}}

\newcommand{\G}{\Gamma}
\newcommand{\cF}{{\cal F}}
\newcommand{\cG}{{\cal G}}
\newcommand{\oD}{\overline D}
\renewcommand{\a}{\alpha}
\renewcommand{\b}{\beta}
\renewcommand{\d}{\delta}
\newcommand{\D}{\Delta}
\newcommand{\w}{\vec{w}}
\newcommand{\rmd}{{\rm d}}

\newcommand{\bt}[1]{{\bar t}}
\newcommand{\ts}{\textstyle}

\newcommand{\half}{{\ts \frac{1}{2}}}
\documentclass[11pt]{article}
\input epsf
\usepackage{amsmath,amsfonts,epsfig,color,latexsym}
\usepackage{amssymb}
\usepackage{curves}
\input feynman
\csname @addtoreset\endcsname{equation}{section}
\begin{document}

\begin{titlepage}
{\hbox to\hsize{\hfill {AEI--2003--001}}} 
{\hbox to\hsize{${~}$ \hfill {LAPTH-960/03}}} 
\vskip 0.2pt \hfill {\tt hep-th/0301058}

\begin{center}
\vglue .3in {\Large \bf  On a Large ${N}$ Degeneracy in ${\cal N}=4$ SYM\\
and the AdS/CFT Correspondence
}

\vskip 1cm

{\large {$\rm{G.~ Arutyunov}^{(a)}$}\footnote[1]{On leave of absence from Steklov Mathematical Institute, Gubkin str. 8,
117966, Moscow, Russia}
, $\rm{E.~Sokatchev^{(b)}}$  }
\\[.3in]
\small

$^{\rm(a)}$ {\it Max-Planck-Institut f\"ur Gravitationsphysik,
Albert-Einstein-Institut, \\
Am M\"uhlenberg 1, D-14476 Golm, Germany}
\\
[.03in] $^{\rm(b)}${\it Laboratoire d'Annecy-le-Vieux de Physique 
Th\'{e}orique\footnote[2]{UMR 5108 associ{\'e}e {\`a}
 l'Universit{\'e} de Savoie}
LAPTH, B.P. 110, \\ F-74941 Annecy-le-Vieux et l'Universit\'{e} de Savoie}
\\[.3in]
\normalsize

{\bf ABSTRACT}\\[.0015in]
\end{center}
We study the four-point correlator of $\half$-BPS operators of weight 4 in
${\cal N}=4$ SYM, which are dual to massive KK modes in AdS${}_5$ supergravity.
General field-theoretic arguments lead to a partially non-renormalized form of the amplitude that depends on two {\it a priori} independent functions of the conformal cross-ratios. We explicitly compute the amplitude in the large $N$ limit at one loop (order $g^2$) and in AdS${}_5$ supergravity. \\
Surprisingly, the one-loop result shows that the two functions determining the amplitude {\it coincide} while in the supergravity regime they are  {\it distinctly different}. We discuss the possible implications of this perturbative degeneracy for the AdS/CFT correspondence.

\vskip 3cm
\vskip 7pt ${~~~}$ \newline

\end{titlepage}

\section{Introduction}  
The holographic duality provides a fascinating relationship between gauge theories and strings. In particular, a lot of progress has recently been made in understanding the holographically dual pair formed by the (strongly-coupled) ${\cal N}=4$ supersymmetric Yang-Mills theory (SYM) and type IIB (supergravity) superstring on an AdS background. Still, our confidence in this example is primarily based upon the fact that both theories exhibit the same powerful superconformal symmetry. It is therefore highly desirable to find a way to analyze and compare, even qualitatively, their true dynamical features. 

The compactification of type IIB supergravity on an $AdS_5\times S^5$ background results in an infinite tower of massive Kaluza-Klein (KK) modes. In the dual gauge theory they correspond to $\half$-BPS protected multiplets. These multiplets are rather special (short) as their lowest weight states are annihilated by half of the Poincar\'e supercharges. Hence, in the quest for common dynamical features one can try, in particular, to compute the correlation functions of the $\half$-BPS operators both perturbatively and in the supergravity regime, and then to compare them. This concerns in the first place the four-point correlators which, unlike the two- and three-point functions receive quantum corrections. Certainly, we do not expect to find a literal agreement since, by the logic of the AdS/CFT duality, the two regimes correspond to small and infinite values of the 't Hooft coupling $\lambda=g^2N$ ($N$ is the rank of the gauge group), respectively. It is known that the form of the four-point amplitude is partially fixed by the superconformal Ward identities (together with crossing symmetry). These are purely kinematical restrictions and of course they still leave a substantial functional freedom to account for the non-trivial dynamics. It is possible, however, to further reduce this freedom by using the well-known field-theoretic insertion procedure. The new constraints follow from the fact that the quantum corrections are generated by inserting the SYM action into the amplitude. Such results go beyond the pure kinematics as they essentially depend on the Lagrangian description of the gauge theory. Then, what we can test by comparing the supergravity-induced and the field-theory (perturbative or instanton) amplitudes is their {\it partial non-renormalization}, i.e. the dynamically constrained form of the amplitude in comparison with the general solution of the superconformal Ward identities.

There exists a general procedure for determining the maximal number of independent functions of the conformal cross-ratios in the four-point correlator of $\half$-BPS operators with arbitrary weights (dimensions) $k_p$, $p=1,2,3,4$. The insertion of the SYM action effectively reduces the weight at each point by two units. The resulting object of lower weight depends on as many functions as allowed by its crossing and R symmetry properties. In particular, if all the $k_p$ equal 2 or 3, the quantum part of the correlator should depend on a single function of the conformal cross-ratios. The supergravity-induced four-point amplitudes for $k_p=2$ and 3 are now available. Remarkably, these amplitudes split into a ``free'' and an ``interacting'' parts; the latter is determined by a single function of the cross-ratios, in precise agreement with the field theory prediction. We recall that the case $k_p=3$ is the first non-trivial example of a $\half$-BPS operator dual to a massive KK mode. These results are rather reassuring and support the conviction that the four-point correlators of $\half$-BPS operators of arbitrary weight derived from the effective supergravity Lagrangian obey the corresponding partial non-renormalization theorems.    

In this paper we continue the analysis of the correlation functions of $\half$-BPS operators dual to the higher KK modes of the supergravity theory. The next example to consider is $k_p=4$. Our motivation to undertake this study is not only to confirm, once more, the partial non-renormalization. We are interested in this case for the following two main reasons. 

The first example of a {\it generic} $\half$-BPS multiplet corresponds to $k=4$. Indeed, the multiplet with $k=2$ is rather special (``ultrashort") as it contains the conserved stress tensor and R currents of the theory. Its dual is the graviton multiplet of gauged ${\cal N}=8$ supergravity which comprises the massless KK modes of the compactified ten-dimensional theory. The $k=3$ multiplet also exhibits some exceptional shortening compared to multiplets with $k\geq 4$. The generic nature of the $\half$-BPS multiplet with  $k=4$ is also reflected in the structure of the cubic effective Lagrangian of $AdS_5$ supergravity: From $k=4$ on, several new scalar and vector fields mediating the interactions of the KK scalars $s_k$ emerge. 

The insertion procedure predicts that the ``quantum part'' of the four-point amplitude of the $\half$-BPS operators of weight 4 involves two {\it a priori} independent functions $\cF$ and $\cG$ with different crossing symmetry properties:
\bea
\nonumber
{\cal F}(s/t,1/t)=t\;{\cal F}(s,t) \, , ~~~~~~~~~{\cal G}(1/s,t/s)=s\;{\cal G}(s,t)\, ,
\eea
where $s,t$ are the conformal cross-ratios. Our second motivation is to find out if there exist some unexpected relations between these two functions. It is crucial to realize that such relations can only be due to some new, dynamical mechanism. 

In an attempt to investigate these issues, in the present paper we perform two distinct computations. Firstly, we calculate the one-loop (order $\lambda$) four-point amplitude for $\half$-BPS operators of weights $k=2,3,4$. Secondly, using the AdS supergravity effective action we derive the supergravity-induced four-point amplitude for operators with $k=4$. We then show that in both cases the corresponding four-point amplitudes do exhibit the expected partial non-renormalization. 

What comes out as a surprise is that in the large $N$ limit the two functions $\cF$ and $\cG$ {\it coincide} at one loop, while in the supergravity regime they are {\it distinctly different}! Thus, compared to the supergravity result, which according to the AdS/CFT duality should match  the large $N$ limit of the gauge theory,  the one-loop amplitude exhibits a {\it degenerate behavior}: A single function is sufficient to describe the correlator.\footnote{For the four-point amplitude of single-trace $\half$-BPS operators this degeneracy is a large $N$ effect and it is lifted as soon as $1/N$ corrections are taken into account. However, allowing mixing with double-trace operators, it might be possible to restore the degeneracy even for finite $N$.} It is therefore urgent to understand how the higher-loop (at least, the two-loop) or the instanton  corrections affect this degeneracy. If it persists, we would be facing a real puzzle in the context of the AdS/CFT correspondence. 

Another interesting feature of the one-loop result is that for weight $k$ the function $\cF_k \propto k^2\Phi(s,t)$, where $\Phi(s,t)$ is the one-loop scalar box. This universal dependence on the weight has been verified by explicit computations for $k=2,3,4$. It is interesting to find out whether for arbitrary $k$ the four-point correlator is still described by the same single function $\cF_k$. 
   
The paper is organized as follows. In Section 2 we describe the general form of the four-point amplitude for $\half$-BPS operators of weight 4 determined by its conformal, R and crossing symmetries. It depends on four arbitrary functions of the conformal cross-ratios. Then we recall how the insertion of the SYM action reduces the number of independent functions in the ``quantum'' part of the amplitude from four to two. In Section 3, working in the ${\cal N}=2$ harmonic superspace approach, we again make use of the insertion procedure to compute the one-loop amplitude of $\half$-BPS operators of weights 2, 3 and 4. Section 4 is devoted to the supergravity analysis, where we compute the supergravity-induced amplitude for the weight 4 operators and explicitly identify the functions $\cF$ and $\cG$. Finally, in Section 5 we summarize our perturbative and supergravity findings for $k=2,3,4$. The computational details are gathered in three Appendices.

\section{Generalities}\label{general}

Here we summarize some basic facts (following Ref. \cite{ADOS}) about the four-point correlators of $\half$-BPS operators and fix our notation.

We consider $\half$-BPS operators of conformal weight $k$ realized as ${\cal N}=4$ single-trace composite operators ${\cal O}^I$ with a suitably normalized 
two-point function: 
\bea \label{ibasis} {\cal O}^I= 
C_{i_1\cdots i_k}^I \mbox{Tr}(\phi^{i_1} \cdots \phi^{i_k}) \, . \eea 
Here $\phi^i$, $i=1,\ldots, 6$ are the ${\cal N}=4$ SYM scalars and 
$C^I_{i_1\cdots i_k}$ are traceless symmetric tensors obeying the 
normalization condition $C^I_{i_1\cdots i_k}C^J_{i_1\cdots 
i_k}=\d^{IJ}$, which describe the irreducible representations $[0,k,0]$. 
We want to study the four-point correlator $\langle {\cal O}^{I_1}{\cal O}^{I_2}{\cal O}^{I_3}{\cal O}^{I_4}  \rangle \equiv \langle {\cal O}^{1}{\cal O}^{2}{\cal O}^{3}{\cal O}^{4}  \rangle$. General considerations based on conformal covariance and on the R symmetry SO(6) imply the following expression 
for the four-point amplitude in the case $k=4$: 
\bea
\nonumber
\langle {\cal O}^{1}{\cal O}^{2}{\cal O}^{3}{\cal O}^{4}  \rangle &=&
a_1\frac{\d^{12}\d^{34}}{x_{12}^8x_{34}^8}+a_2\frac{\d^{13}\d^{24}}{x_{13}^8x_{24}^8}
+a_3\frac{\d^{14}\d^{23}}{x_{14}^8x_{23}^8}\\
\nonumber
&+&b_1\frac{C^{1234}}{x_{12}^6x_{34}^6x_{13}^2x_{24}^2}
+b_2\frac{C^{1243}}{x_{12}^6x_{34}^6x_{14}^2x_{23}^2}
+b_3\frac{C^{1342}}{x_{13}^6x_{24}^6x_{14}^2x_{23}^2}
\\
\label{Ampl}
&+&b_4\frac{C^{1324}}{x_{13}^6x_{24}^6x_{12}^2x_{34}^2}
+b_5\frac{C^{1423}}{x_{14}^6x_{23}^6x_{12}^2x_{34}^2}
+b_6\frac{C^{1432}}{x_{14}^6x_{23}^6x_{13}^2x_{24}^2}
\\
\nonumber
&+&c_1\frac{\Omega^{1234}}{x_{12}^4x_{13}^4x_{24}^4x_{34}^4}
+c_2\frac{\Omega^{1243}}{x_{12}^4x_{14}^4x_{23}^4x_{34}^4}
+c_3\frac{\Omega^{1432}}{x_{13}^4x_{14}^4x_{23}^4x_{24}^4}\\
\nonumber
&+&
d_1\frac{\Upsilon^{1234}}{x_{12}^4x_{34}^4x_{13}^2x_{14}^2x_{23}^2x_{24}^2}
+d_2\frac{\Upsilon^{1324}}{x_{13}^4x_{24}^4x_{12}^2x_{14}^2x_{23}^2x_{34}^2}
+d_3\frac{\Upsilon^{1432}}{x_{14}^4x_{23}^4x_{12}^2x_{13}^2x_{24}^2x_{34}^2}
\eea
Here the 15 coefficients $a,b,c$ and $d$ are functions of the two conformal cross-ratios
\bea
s=\frac{x^2_{12}x^2_{34}}{x^2_{13}x^2_{24}}\,, \qquad
t=\frac{x^2_{14}x^2_{23}}{x^2_{13}x^2_{24}} \, .
\eea
To keep track of the R symmetry structure of the four-point amplitude 
we find it convenient 
to introduce the following $SO(6)$-invariant tensors  
\bea
\nonumber
\d^{12}\d^{34}&=&C_{ijkl}^{1}C_{ijkl}^{2}C_{mnsp}^{3}C_{mnsp}^{4} \, ,\\
\nonumber
C^{1234}&=&C_{ijkl}^{1}C_{ijkm}^{2}C_{nspl}^{3}C_{nspm}^{4} \, ,\\
\label{propbasis}
\Omega^{1234}&=&C^1_{ijkl}C^2_{ijsp}C^3_{mnkl}C^4_{mnsp}\, , \\
\nonumber
\Upsilon^{1234}&=&C^1_{ijkl}C^2_{ijsp}C^3_{mnks}C^4_{mnlp} \, 
\eea
and permutations thereof. The C-tensors posses the following symmetry
\bea
\nonumber
C^{1234}=C^{2143}=C^{3412}=C^{4321}\, ,
\eea
while the tensors $\Omega$ and $\Upsilon$ in addition to the same permutation symmetry
obey the relations
\bea
\nonumber
\Omega^{1234}=\Omega^{1324}\, , ~~~~\Upsilon^{1234}=\Upsilon^{1243} \, .
\eea 
The four-point amplitude (\ref{Ampl}) contains 15 propagator structures (see Figure 1) which according to Ref. \cite{ADOS} can be grouped together into different classes invariant under crossing symmetry. For $k=4$ every set is described by a triplet of integers $(m,n,l)$ such that  $m\geq n\geq l\geq 0$ and $m+n+l=4$. Obviously, we have four such sets. The explicit crossing symmetry relations among the coefficients of the amplitude within each class are
\begin{itemize}
\item{$(4,0,0)$}
\bea
\nonumber
a_1(s,t)&=&a_3(t,s)=a_1(s/t,1/t) \\
a_2(s,t)&=&a_2(t,s)=a_3(s/t,1/t); 
\eea 
\item{$(3,1,0)$}
\bea
\nonumber
b_1(s,t)&=&b_2(s/t,1/t)=b_4(1/s,t/s) \\
b_2(s,t)&=&b_3(1/s,t/s)=b_5(t,s)=b_6(1/t,s/t);
\eea
\item{$(2,2,0)$}
\bea
\nonumber
c_1(s,t)&=&c_2(s/t,1/t)=c_3(t,s); 
\eea 
\item{$(2,1,1)$}
\bea
\nonumber
d_1(s,t)&=&d_2(1/s,t/s)=d_3(t,s)\, . 
\eea 
\end{itemize}
Thus, modulo crossing symmetry the four-point amplitude depends 
on {\it four independent functions}, which we can choose
to be $a_1,b_1,c_1$ and $d_1$.  

We note also that the 15 propagator structures in eq. (\ref{Ampl})
are related to the 15 channels in the tensor product decomposition 
\bea
\nonumber
[0,4,0]_{105}\times [0,4,0]_{105}&=&[0,0,0]_1+[0,2,0]_{20}+[0,4,0]_{105}+[0,6,0]_{336}+
[0,8,0]_{825}\\
\label{tp}
&&\hspace{-3cm} +[2,0,2]_{84}+[2,2,2]_{729}+[2,4,2]_{2640}+[4,0,4]_{825}\\
\nonumber
&&\hspace{-3cm}
+[1,0,1]_{15}+[1,2,1]_{175}+[1,4,1]_{735}+[1,6,1]_{2079}+[3,0,3]_{300}+[3,2,3]_{2156}\, .
\eea
The subscript indicates the dimension of the corresponding irrep of SO(6). The irreps in the first two lines of eq. (13) are symmetric and those in the third line are antisymmetric in the indices $I_1,I_2$. Therefore the OPE implied by the four-point 
amplitude  (\ref{Ampl}) will have 15 different SO(6) channels. 
According to the classification in Refs. \cite{AES} only six of them may contain unprotected superconformal primary operators: $[0,0,0]$, $[0,2,0]$, $[0,4,0]$, $[2,0,2]$, $[1,0,1]$ and [1,2,1].

Further, dynamical restrictions on the {\it quantum part} of  the four-point amplitude (\ref{Ampl}) follow from the field-theoretic insertion 
formula \cite{ADOS}.\footnote{See also Ref. \cite{HH} for an alternative argument.} Namely, the quantum corrections factorize into a fixed prefactor of weight 2 and an arbitrary factor of weight $k-2$: 
\bea
\frac{\pa}{\pa {g^2}}
\langle {\cal O}^{1}{\cal O}^{2}{\cal O}^{3}{\cal O}^{4}  \rangle &=&
R^{2222}F^{(k-2)}\, .  \label{N4ins}
\eea
Here 
\bea
R^{2222}=\frac{1}{x_{13}^2x_{24}^2}\Biggl[
\frac{(12)^2(34)^2}{x_{12}^2x_{34}^2}+
\frac{(13)(14)(23)(24)}{x_{13}^2x_{14}^2x_{23}^2x_{24}^2}
(x_{12}^2x_{34}-x_{14}^2x_{23}^2-x_{13}^2x_{24}^2)+ \mbox{cycle}
\Biggr] \label{r2222}
\eea
is the weight 2 prefactor independent of the value of $k$. The symbols $(12)^2(34)^2$, etc. stand for SO(6) harmonic-projected propagator structures (see Section \ref{puresect}). The remaining dynamical information is encoded in the function $F^{(k-2)}$ of weight $k-2$ at each point. In our particular case $k=4$ this function is
 \bea
F^{(2)}&=&\frac{(12)^2(34)^2}{x_{12}^4x_{34}^4}\a_1
+\frac{(13)^2(24)^2}{x_{13}^4x_{24}^4}\a_2
+\frac{(14)^2(23)^2}{x_{14}^4x_{23}^4}\a_3\\
\nonumber
&+&\frac{(13)(14)(23)(24)}{x_{13}^2x_{14}^2x_{23}^2x_{24}^2}\b_1
+\frac{(12)(14)(23)(34)}{x_{12}^2x_{14}^2x_{23}^2x_{34}^2}\b_2
+\frac{(12)(13)(24)(34)}{x_{12}^2x_{13}^2x_{24}^2x_{34}^2}\b_3 \, ,
\eea
where $\a_p(s,t)$ and $\b_p(s,t)$ are some unknown functions of the conformal cross-ratios. 

\begin{figure}[tbp]
\begin{center}
\input{fig4.pstex_t}
\end{center}\caption{Propagator structures for the case $k=4$.}\label{k4pro}
\end{figure}
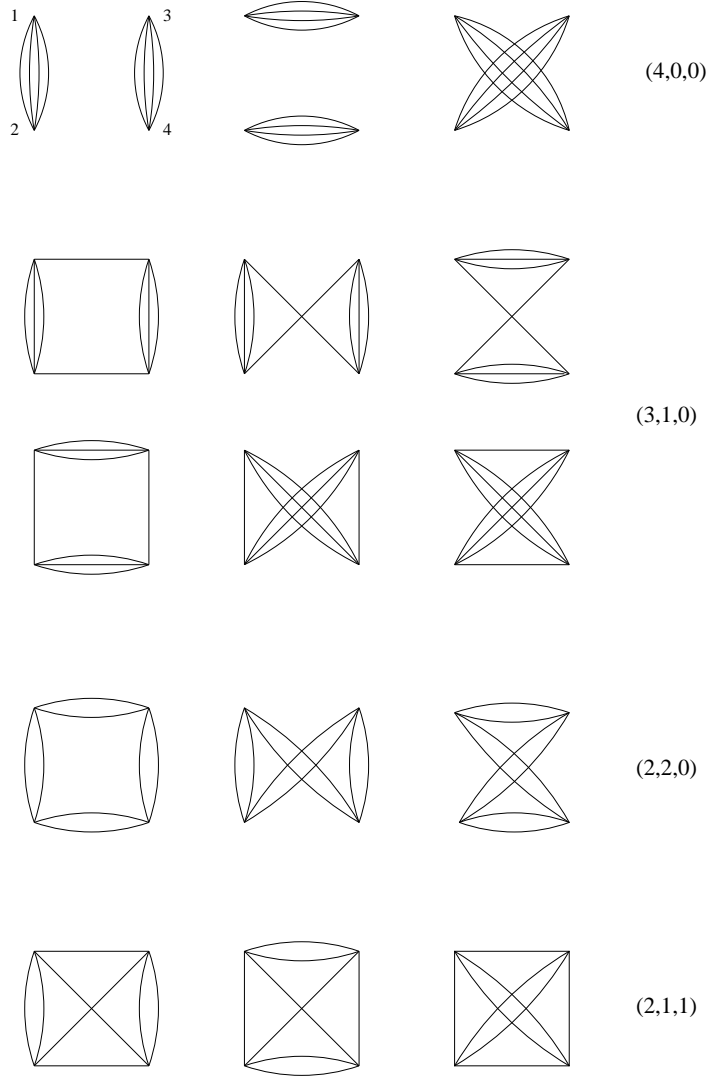

This factorized form implies further restrictions on the 
coefficients of the four-point amplitude (\ref{Ampl}). Indeed, expanding the product of the prefactor $R^{2222}$ with the function $F^{(2)}$ and matching the 
propagator structures arising with those in eq. (\ref{Ampl}), we can  
express  the original coefficients $a,b,c,d$ in terms of 
$\a_p$ and $\beta_p$. The coefficients $a_p$ are given by  
\bea
\label{a}
a_1(s,t)=s\a_1(s,t)\, ,~~~~~a_2(s,t)=\a_2(s,t)\, , ~~~~~a_3(s,t)=t\a_3(s,t) \, .
\eea
For the coefficients $b_p$ we obtain 
\bea
\nonumber
b_1&=&s\b_3+\a_1(t-s-1)\, ,  ~~~~~~~~~b_4=\b_3+\a_2(t-s-1)\, , \\
\label{b}
b_2&=&s\b_2+\a_1(1-s-t)\, ,  ~~~~~~~~~b_5=t\b_2+\a_3(1-s-t) \, ,\\
\nonumber
b_3&=&\b_1+\a_2(s-t-1) \, ,  ~~~~~~~~~~b_6=t\b_1+\a_3(s-t-1)  \, .
\eea
For $c_p$ we have
\bea
\nonumber
c_1&=&\a_1+s\a_2+(t-s-1)\b_3 \, ,\\
\label{c}
c_2&=&t\a_1+s\a_3+(1-s-t)\b_2 \, ,\\
\nonumber
c_3&=&\a_3+t\a_2+(s-t-1)\b_1 \, .
\eea
Finally, the coefficients $d_p$ read  
\bea
\nonumber
d_1&=&\a_1(s-t-1)+s\b_1+\b_2(t-s-1)+\b_3(1-t-s) \, ,\\
\label{d}
d_2&=&\a_2(1-s-t)+\b_2+\b_1(t-s-1)+\b_3(s-t-1) \, ,\\
\nonumber
d_3&=&\a_3(t-s-1)+t\b_3+\b_1(1-s-t)+\b_2(s-t-1) \, . 
\eea

Now, under point permutations the prefactor $R^{2222}$ transforms as follows 
\begin{eqnarray}
1\leftrightarrow 2: && R^{2222} \ \rightarrow \ \frac{1}{t}\; R^{2222} \nonumber\\
1\leftrightarrow 3: && R^{2222} \ \rightarrow \  R^{2222}\;  \label{crossR}
\end{eqnarray}
(the harmonic factors are symmetric, e.g. $(12)=(21)$).
This implies that crossing symmetry relates the three coefficients $\a_p$ as follows: \bea \label{cross1}
\a_1(s,t)=\a_3(t,s)=1/s~\a_2(t/s,1/s)=1/t~\a_1(s/t,1/t) \, ,
\eea
while for the $\b$'s we obtain 
\bea   \label{cross2}
\b_3(s,t)=\b_1(t,s)=1/t~\b_2(s/t,1/t)=1/s~\beta_3(1/s,t/s)\, .
\eea
Thus, the insertion formula reduces the number of independent functions from four to two. This is the content of the partial renormalization theorem for $\half$-BPS operators of weight 4. We identify these two independent functions with, e.g., ${\cal F}\equiv \a_1$ and ${\cal G}\equiv \b_3$, satisfying a single crossing symmetry 
condition each:
\bea
\label{cr}
{\cal F}(s,t)=1/t~{\cal F}(s/t,1/t) \, , ~~~~~{\cal G}(s,t)=1/s~{\cal G}(1/s,t/s)\, .
\eea 
In Section \ref{oneloopamp} we compute these functions in perturbation theory at one loop, and in Section \ref{gravity} in the supergravity approximation, thus confirming the partial non-renormalization theorem.

We conclude this section by presenting the free-field theory values of the  coefficient functions in the large $N$ limit:
\bea
\label{free}
a_{1,2,3}=1\, , ~~~~~~b_{1,\cdots,6}=\frac{16}{N^2}\, , ~~~~~~c_{1,2,3}=\frac{16}{N^2}\, , ~~~~~~
d_{1,2,3}=\frac{32}{N^2} \, .
\eea  

\section{One-loop four-point amplitudes} \label{oneloopamp}

In this section we compute the amplitude (\ref{Ampl}) at one loop using ${\cal N}=2$ Feynman rules in harmonic superspace \cite{HSSFR}. This technique has two advantages. Firstly, the calculation is very simple, being reduced to just a single graph. Secondly, in an ${\cal N}=2$ setup we still see a non-trivial part of the initial R symmetry group, SU(2) $\subset$ SU(4), which allows us to directly identify the various coefficients in the amplitude (\ref{Ampl}). 

\subsection{Reducing ${\cal N}=4$ to ${\cal N}=2$. ``Pure" projections} \label{puresect}

We want to compute the four-point correlator of $\half$-BPS operators of weight $k$  
\begin{equation}\label{1}
  \langle {\cal O}^{(k)}{\cal O}^{(k)}{\cal O}^{(k)}{\cal O}^{(k)} \rangle|_{\theta=0}
\end{equation}
using ${\cal N}=2$ Feynman diagrams. Here ${\cal O}^{(k)} = {\rm Tr}({\cal W }^{\{i_1}\cdots {\cal W }^{i_k\}})$; ${\cal W}^i$, $i=1,\ldots,6$ is the ${\cal N}=4$ field-strength superfield; $\{\}$ denotes traceless symmetrization.
To this end we first have to decompose each ${\cal O}^{(k)}$ into its ${\cal N}=2$ hypermultiplet (HM) and SYM constituents. As we show in this subsection (following \cite{EHSSW,ADOS}), for $k\leq 4$ the complete  ${\cal N}=4$ four-point correlator can be reconstructed just from one HM projection  of the simplest, ``pure" type. 

The lowest component of the ${\cal N}=4$ field-strength multiplet $\phi^i(x) = {\cal W}^{i}|_{\theta=0}$ is a real vector of SO(6). Reducing SO(6) to SU(3), we 
can decompose it into $3+\bar 3$:
\begin{equation}\label{6}
  \phi^{i} \ \rightarrow \ \phi^{\bf{i}}, \ \bar\phi_{\bf{i}}\;, \quad {\bf{i}}=1,2,3\;.
\end{equation}
The further decomposition SU(3) $\rightarrow$ SU(2)$\times$U(1) results 
in
\begin{equation}\label{7'}
   \phi^{\bf{i}} \ \rightarrow \ \phi^A \equiv \varphi^A,\ A=1,2; \quad \ \phi^3 \equiv w\;.
\end{equation}
After projection with SU(2)/U(1) harmonics,
\begin{equation}\label{su2harm}
  u^\pm_A \in \mbox{SU(2)}: \qquad u^+_A \epsilon^{AB} u^-_B = 1, \ \ \overline{u^{+A}} = u^-_A \;,
\end{equation}
the field $\varphi^A(x)$ becomes the lowest 
component of the on-shell ${\cal N}=2$ Grassmann analytic HM superfield  \cite{HSSFR}
\begin{equation}\label{qplus}
  q^+(x,\theta^+,\bar\theta^+,u^\pm) =  
\varphi^A(x)u^+_A + \theta^{+\alpha}\psi_\alpha(x) + \bar\theta^+_{\dot\alpha} \bar\kappa^{\dot\alpha}(x)+ 2i \theta^+\sigma^\mu\bar\theta^+ \pa_\mu \phi^A(x)u^-_A \,,
\end{equation}
where $\theta^+_\alpha = u^+_A \theta^A_\alpha$, $\bar\theta^+_{\dot\alpha} =  u^+_A \bar\theta^A_{\dot\alpha}$ and $\psi_\alpha(x),\bar\kappa^{\dot\alpha}(x)$ are the fermions in the HM. Further, the field $w(x)$ becomes the lowest component of the chiral ${\cal N}=2$ field 
strength $W(x,\theta) = w(x) + \theta^A_\alpha \lambda^\alpha_A(x) + i\theta^A\sigma_{\mu\nu}\theta_A F^{\mu\nu}(x) $. The HM superfield $q^+$ has a harmonic superspace conjugate which is also Grassmann analytic, $\tilde q^+(x,\theta^+,\bar\theta^+,u^\pm)  = u^{+A} \bar\varphi_A(x) + \cdots$, while the conjugate of the chiral filed strength $W(x,\theta)$ is the antichiral $\bar W(x,\bar\theta)$.

Like in the case of SU(2), a convenient way of keeping track of the SO(6) indices is to project them with harmonic variables. Now this is a complex vector $z_i$ satisfying the conditions 
\begin{equation}  
z_iz_i = 0\;, \qquad z_i\bar{z}_i = 1\;. 
\label{defhar}
\end{equation}
This vector provides a harmonic description of the coset space 
SO(6)/SO(2)$\times$SO(4). With its help we can project ${\rm Tr}({\cal W }^{\{i_1}\cdots {\cal W }^{i_k\}})$ onto the highest-weight state of the representation $[0,k,0]$: 
\begin{equation}\label{prO} 
{\cal W}^{k} = z_{i_1} \cdots z_{i_k} \mbox{Tr}({\cal W}^{i_1} \cdots 
{\cal W}^{i_k}) \, . 
\end{equation}
Here the Dynkin label $k$ is identified with the 
U(1) charge of the projection (\ref{prO}) (assuming that the vector 
$z_i$ carries U(1) charge $+1$). 

The free propagator (two-point function) for the 
elementary ${\cal N}=4$ SYM scalars is 
\begin{equation}\label{frpr} 
\langle {\cal W}^i(1){\cal W}^j(2) \rangle|_{\theta=0}  = \langle \phi^i(x_1)\phi^j(x_2) \rangle = \frac{\delta^{ij}}{x^2_{12}} \;. 
\end{equation}                             
Using two copies of the SO(6) harmonic variables, one for each point, we can project the propagator (\ref{frpr}): 
\begin{equation}\label{prpr} 
\langle {\cal W} (1){\cal W} (2) \rangle|_{\theta=0} =\langle \phi(1)\phi(2) \rangle = \frac{z_{1i} \delta^{ij} z_{2j}}{x^2_{12}} \equiv \frac{(12)}{x^2_{12}} = 
\frac{(21)}{x^2_{21}}\;. 
\end{equation}       

Next, reducing SO(6) to SU(3) we decompose the contraction of the SO(6) harmonics into SU(3) pieces:
\begin{equation}\label{9'}
 ({12}) = 1_i\delta^{ij}2_j = 1^{\bf i} \bar 2_{\bf i} + \bar 1_{\bf i} 2^{\bf i} \equiv [1\bar 2] 
+ [\bar 12]\;.
\end{equation}
In this notation we have ``oriented" two-point functions for the SU(3)-covariant 
field strengths: $\langle {\cal W}\bar {\cal W} \rangle = [1\bar 2]$ and $\langle \bar{\cal W} {\cal W} \rangle = [\bar 12]$ (the space-time dependence is suppressed; in addition, we always set $\theta=0$, unless stated otherwise). The further reduction of SU(3) to SU(2)$\times$U(1) gives, e.g., $[1\bar 2] = 1^{\bf i} \bar 2_{\bf i} = 1^A \bar 2_A + 1^3 \bar 2_3$. This can be split into two independent propagators, one 
for the ${\cal N}=2$ HM:
\begin{equation}\label{10}
 \langle q\tilde q \rangle \sim  1^A \bar 2_A = 1^A \epsilon_{AB}2^B = 
-\bar 1_A 2^A \equiv [12] = -[21]
\end{equation}
and one for the ${\cal N}=2$ field strength, $\langle W\bar W \rangle 
\sim 1^3 \bar 2_3 \equiv 1$ (in the latter there is no need to use 
harmonics, $1^3 \bar 2_3$ is just a ``bookkeeping device").

Consider now the four-point correlators of the SO(6) harmonic-projected $\half$-BPS operators ${\cal W }^k$.  These four-point functions have a harmonic 
structure consisting of all possible pairings of the four sets of 
harmonics. For instance, for $k=4$ we have:
\begin{eqnarray}
&&\hskip-1cm \langle {\cal W }^4 |{\cal W }^4 |{\cal W }^4|{\cal W }^4\rangle \nonumber\\
&=&{} A_1\;({12})^4({34})^4 + A_2\;({13})^4({24})^4 + A_3\;({14})^4({23})^4   
\nonumber \\
&& {}+  B_1\;({12})^3({34})^3({13})({24}) + B_2\;({12})^3({34})^3({14})({23}) +
B_3\;({13})^3({24})^3({14})({23}) \nonumber\\
&& {}+  B_4\;({13})^3({24})^3({12})({34}) + B_5\;({14})^3({23})^3({12})({34}) +
B_6\;({14})^3({23})^3({13})({24}) \nonumber\\
&& {}+ C_1 (12)^2(13)^2(24)^2(34)^2 + C_2 (12)^2(14)^2(23)^2(34)^2 + C_3 (13)^2(14)^2(23)^2(24)^2  \nonumber\\
&& {}+ D_1 (12)^2(34)^2(13)(14)(23)(24) + D_2 (13)^2(24)^2(12)(14)(23)(34)  \nonumber\\
&& {} + D_3 (14)^2(23)^2(12)(13)(24)(34) \;. \label{0}
\end{eqnarray}
In fact, eq. (\ref{0}) is just eq. (\ref{Ampl}) rewritten in SO(6) harmonic notation. Here we have absorbed the space-time propagator factors into the coefficient functions $A,B,C,D$, e.g., $A_1 = a_1/(x^8_{12}x^8_{34})$. 

The reduction to either ${\cal N}=2$ HMs or SYM field strengths is 
straightforward. We replace each SO(6) contraction by SU(3) 
contractions, $(pq)= [p\bar q] + [\bar pq]$, and then expand each SO(6) 
harmonic structure in eq. (\ref{0}) into products of SU(3) contractions. If we want to keep only the HM constituents of the composite operators, we need to replace the SU(3) contractions by SU(2) ones respecting the signs, e.g., $[1\bar 2]\rightarrow [12]$, $[\bar 12]\rightarrow -[12]$. Take, for example, the two-point function $\langle {\cal W}^2|{\cal W}^2\rangle \sim (12)^2$. The reduction to SU(3) gives $(12)^2 = [1\bar 2]^2 + 2[1\bar 2][\bar 12] + [\bar 12]^2$. This ${\cal N}=4$ two-point function has two inequivalent HM projections: the ``pure" projection $\langle\tilde q^2|  q^2\rangle \ \rightarrow\ [\bar 12]^2 \ \rightarrow\ [12]^2$ and the ``mixed" projection $\langle\tilde qq| \tilde qq\rangle \ \rightarrow\ 2[1\bar 2][\bar 12] \ \rightarrow\ -2[12]^2$. Generalizing this two-point example, we can easily work out the simplest, ``pure" four-point projection in which we take only $q$'s or only $\tilde q$'s  at each point:
\begin{eqnarray}
 \langle \tilde q^4|q^4|q^4 |\tilde q^4 \rangle &=& 
a_1\;\left[ \frac{[12][43]}{x^2_{12}x^2_{43}}\right]^{4}  + a_2\;\left[ \frac{[13][42]}{x^2_{13}x^2_{42}} \right]^{4} +  b_1\; \left[ \frac{[12][43]}{x^2_{12}x^2_{43}}\right]^{3} \left[ \frac{[13][42]}{x^2_{13}x^2_{42}}\right] \nonumber \\
  && {}  + b_4\; \left[ \frac{[12][43]}{x^2_{12}x^2_{43}}\right]\left[ \frac{[13][42]}{x^2_{13}x^2_{42}}\right]^{3}  + c_1\; \left[ \frac{[12][43]}{x^2_{12}x^2_{43}}\right]^{2} \left[ \frac{[13][42]}{x^2_{13}x^2_{42}}\right]^{2}\;. \label{1'} 
\end{eqnarray}
We see that the pure projection (\ref{1'}) involves all the graphs in Fig. 1 {\it without diagonals}. This is also true for any $k$. So, in general some crossing-equivalence classes are not represented in the pure projection. For $k=4$ this is the case of the last class  $(2,1,1)$ which corresponds to the coefficient functions $d_{1,2,3}$. Still, according to the discussion in Section \ref{general}, the information contained in eq. (\ref{1'}) is sufficient to reconstruct all the 15 coefficients in the ${\cal N}=4$ amplitude; to this end it is enough to know the values of, e.g., $a_1$ and $b_1$.   

Similarly, to obtain the U(1) or chiral-antichiral ${\cal N}=2$ 
field-strength projection we replace every $(pq)$ in eq. (\ref{0}) by 1 if it corresponds to a Wick contraction of the type $\langle \bar W W 
\rangle = \langle W \bar W \rangle$, or by 0 if it corresponds to $\langle WW \rangle$ or to $\langle \bar W \bar W \rangle$. In this way we find
\begin{equation}\label{8''}
  \langle \bar W^4| W^4|W^4|\bar W^4 \rangle =  \frac{1}{x^8_{13}x^8_{42}} \left[a_2 +  \frac{b_4}{s}  + \frac{c_1}{s^2}   + \frac{b_1}{s^3} + \frac{a_1}{s^4}   \right] \;,
\end{equation} 
so the pure U(1) projection involves just a combination of the coefficients appearing in the pure HM projection (\ref{1'}) (this holds for any $k$ as well). 
Finally, by using eqs. (\ref{a}), (\ref{b}), (\ref{c}) and (\ref{cross1}) we obtain
\begin{equation}\label{8''d}
  \langle \bar W^4| W^4|W^4|\bar W^4 \rangle =  \frac{1}{x^8_{13}x^8_{42}} 
 \frac{t}{s}\left[\frac{\cF(1/t,s/t)}{t} + \frac{\cG(s,t)}{s}   + \frac{\cF(s,t)}{s^2} \right] \;.
\end{equation}       
It is clear that from the U(1) projection (\ref{8''d}) one cannot read off the individual coefficients of the ${\cal N}=4$ amplitude. In contrast, the pure HM projection allows us to do this up to $k=4$ (for higher values of $k$ we would have to consider some of the mixed projections too).

\subsection{The ${\cal N}=2$ insertion procedure}\label{n2inspr}

We are interested in the quantum corrections to the lowest component (at $\theta^+_{1,2,3,4}=0$) of the four-point HM correlator $\langle \tilde q^k|q^k|q^k |\tilde q^k\rangle$. The most efficient way to compute them is to employ the ${\cal N}=2$ insertion procedure \cite{hssw}-\cite{ESSun}. Apart from the concrete perturbative calculation, this procedure can also be used to justify the special form (\ref{N4ins}) of the ${\cal N}=4$ amplitude (see Refs. \cite{EPSS,ADOS}).

The quantum corrections to the correlator $\langle \tilde q^k|q^k|q^k |\tilde q^k\rangle$ can be obtained by inserting the  ${\cal N}=2$ SYM action 
\begin{equation}\label{SYMact}
  S_{\mbox{\scriptsize N=2 SYM}} =
\int d^4xd^4\theta\; {\cal L} \;, \qquad {\cal L} = \frac{1}{4g^2}  {\rm Tr}\;W^2  \end{equation}  into the four-point correlator:
\begin{equation}\label{ins}
  \frac{\pa}{\pa g^2}\langle \tilde q^k|q^k|q^k |\tilde q^k\rangle|_{\theta^+_{1,2,3,4}=0}  \propto \int d^4x_0d^4\theta_0\; \langle {\cal L}(x_0,\theta_0) \tilde q^k(1) q^k(2) q^k(3) \tilde q^k(4) \rangle|_{\theta^+_{1,2,3,4}=0}  
\end{equation}
(see Refs. \cite{I,hssw} for a discussion of this standard field-theoretic procedure in the present context). The five-point  ${\cal N}=2$ superconformal covariant under the integral is nilpotent and has the following general form (to the lowest order in the $\theta$ expansion):
\begin{equation}\label{4}
  \langle {\cal L}(0) \tilde q^k(1) q^k(2) q^k(3) \tilde q^k(4) \rangle = \Theta^{2222} \ F^{(k-2)} (x,u) + O(\theta\bar\theta)\;.
\end{equation}
Here $\Theta^{2222}$ is a fixed nilpotent prefactor (see below) and  $F^{(k-2)}(x,u)$ is a conformally covariant function of charges $k-2$ at each point, whose determination is the aim of our one-loop calculation. 

The form (\ref{4}) is a consequence of the superconformal properties of the  five-point correlator. First of all, it has R charge $+4$ (in units in which a left-handed $\theta_\alpha$ has charge $+1$) coming from the ${\cal N}=2$ SYM Lagrangian ${\cal L} \sim W^2$ (the HMs have no R charge). Since the only superspace coordinates with non-vanishing R charge are the $\theta$'s, it is clear that the $\theta$ expansion of this correlator must start with 4 left-handed $\theta$'s. This explains why the five-point function is nilpotent. Further, it is a superconformal invariant (to the lowest order in the $\theta$ expansion, otherwise it is a covariant). A simple counting argument tells us that we can build precisely 4 combinations of the left-handed $\theta$'s invariant under conformal supersymmetry (to lowest order). They are obtained in two steps: (i) first we form 8  combinations of the 4 chiral Grassmann variables $(\theta_0)^A_\alpha$ at the insertion point and of the 8 left-handed  Grassmann analytic variables $(\theta^+_p)_\alpha$, $p=1,2,3,4$ at the HM points, which are invariant under the 4 left-handed $Q$ supersymmetries $\delta_Q \theta_\alpha = \epsilon_\alpha$:
\begin{eqnarray}
\rho_p^{\dot\alpha} =
(\theta_{0}^A u^+_{p\; A} - \theta^+_{p})_\alpha\; \frac{x^{\alpha\dot\alpha}_{0p}}{x_{0p}^2}\;, \qquad p=1,2,3,4\;;
\label{defrho}
\end{eqnarray}     
(ii) next we form 4 combinations of the 8 $\rho$'s invariant under the 4 right-handed $S$ supersymmetries $\delta_S \theta_\alpha = x_{\alpha\dot\alpha} \bar\eta^{\dot\alpha}$, for example,
\begin{equation}\label{4xi}
  \xi^{\dot\alpha}_{12q} = [12]\rho_q^{\dot\alpha} + [2q] \rho_1^{\dot\alpha} + [q1] \rho_2^{\dot\alpha}\;, \qquad q=3,4\;. 
\end{equation}
The invariance of the $\xi$'s can be checked with the help of the harmonic
cyclic identity, e.g.,
\begin{eqnarray}
  [12]3_A + [23]1_A + [31]2_A = 0\;.
\label{harcyc}
\end{eqnarray}
                
Finally, the five-point nilpotent factor with the required R charge is obtained \cite{hssw,ESS,ESSun} by multiplying together all the $\xi$'s (\ref{4xi}):
\begin{eqnarray}
\Theta^{2222} &=& \frac{\xi^2_{123} \xi^2_{124}}{[12]^2} \label{xi2xi2explicit}\\
 &=& \biggl\lbrace [42]^2\rho_1^2\rho_3^2 +2[14][43]\rho_2^2(\rho_1\rho_3) +2[12][23]\rho_4^2(\rho_1\rho_3) + \mbox{cycle in 1,2,3,4}\biggr\rbrace \nonumber\\&&
{}+ \frac{4}{3}\biggl\lbrace
\Bigl[[23][41]+[12][34]\Bigr]
(\rho_1\rho_3)(\rho_2\rho_4)
+\Bigl[[31][24]+[34][21]\Bigr]
(\rho_2\rho_3)(\rho_1\rho_4)
\nonumber\\&&
\qquad
+\Bigl[[13][24]+[14][23]\Bigr]
(\rho_1\rho_2)(\rho_3\rho_4)
 \biggr\rbrace\;,
\nonumber 
\end{eqnarray}   
where $\rho_p\rho_q\equiv \rho^{\dot\alpha}_{p}\epsilon_{\dot\alpha\dot\beta}\rho^{\dot\beta}_q$. In (\ref{xi2xi2explicit}) we have cancelled the overall harmonic factor $[12]^2$ in order to have equal harmonic U(1) charges $+2$ at points 1 to 4.

In the insertion formula (\ref{ins}) we only need the nilpotent covariant (\ref{xi2xi2explicit}) at $\theta^+_{1,2,3,4}=0$. In this frame it depends only on $\theta_0$: 
\begin{eqnarray} &&\hskip-1cm \Theta^{2222} \mid_{\theta^+_{1,2,3,4}=0} =
\frac{\theta_0^4}{x_{01}^2x_{02}^2x_{03}^2x_{04}^2} \label{xichiral}\\
 && \times \Bigl[[13]^2[42]^2 x_{12}^2x_{43}^2 +[12]^2[43]^2x_{13}^2x_{42}^2 + [12][43][13][42]
(x_{14}^2x_{32}^2-x_{13}^2x_{42}^2-x_{12}^2x_{43}^2) \Bigr]\;.
\nonumber \end{eqnarray}
The harmonic polynomial in (\ref{xichiral}) should be compared to $R^{2222}$ in eq. (\ref{r2222}). In fact, the derivation of the ${\cal N}=4$ insertion formula (\ref{N4ins}) goes through its ${\cal N}=2$ version (\ref{xichiral}) (see Appendix A in Ref. \cite{ADOS}).

Another, even simpler form of $\Theta^{2222}$ is obtained by setting, e.g., 
\begin{equation}\label{5}
  \theta_0 = \theta^+_{2,4}=0\ \rightarrow \ \rho_{2,4}=0 \ \rightarrow\  \Theta^{2222} \mid_{\theta_{0,2,4}=0}\; = [42]^2\rho_1^2\rho_3^2\;.
\end{equation}
Since we are only interested in the coefficient function  $F^{(k-2)}(x,u)$, it will be sufficient to calculate the amplitude (\ref{4}) in the frame (\ref{5}). After that we can immediately switch over to the form (\ref{xichiral}) with the explicit $\theta_0$ dependence needed for the chiral integral in (\ref{ins}).

\subsection{Feynman rules}

We want to calculate the one-loop corrections to the four-point correlator. The insertion procedure reduces this to a {\it tree-level calculation} of the five-point function $\langle {\cal L}\tilde q^k q^k q^k \tilde q^k  \rangle$. Consequently, we only need the simplified version of the ${\cal N}=2$ Feynman rules summarized below (see Ref. \cite{hssw,ESS} for details). The lowest component of the HM propagator is shown in Figure \ref{HMpro} ($a,b$ are indices of the adjoint representation of the colour group SU($N$)).
\begin{figure}[ht]
\begin{center}
\hskip-4cm\input{frule1.pstex_t}
\end{center}
\caption{HM propagator}
\label{HMpro}\end{figure}

\begin{figure}[tbp]
\begin{center}
\input{frule2.pstex_t}
\end{center}
\caption{Building blocks}
\label{bblocks}\end{figure}
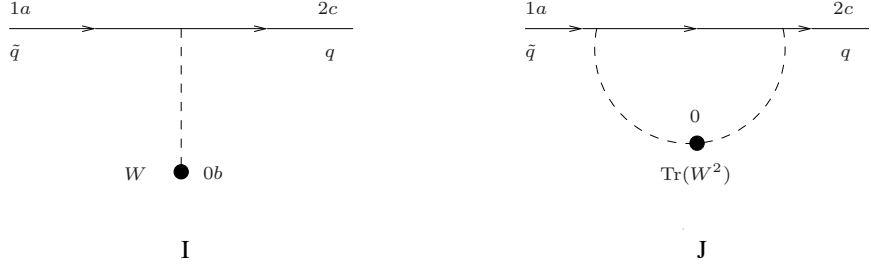

The insertion of the ${\cal N}=2$ SYM Lagrangian gives rise to two distinct building blocks depicted in Fig. \ref{bblocks}. The block $I$ has the expression (to lowest order in the $\theta$ expansion)
\begin{equation}\label{defi12}
  I_{102}^{abc} =  \frac{2gf_{abc}}{(2\pi)^4 x_{12}^2}
\left[[21^-]\rho_1^2+[12^-]\rho_2^2 -2(\rho_1\rho_2)\right]\;.
\end{equation} 
The black blob corresponds to the insertion of a single ${\cal N}=2$ SYM chiral field strength $W$. Consequently, the block (\ref{defi12}) has R charge $+2$. It also carries harmonic U(1) charges $+1$ at points 1 and 2, according to the property of the HMs at these points. 
Notice the appearance of negative charged harmonic variables, e.g., $[12^-] = u^{+}_{1A} \epsilon^{AB} u^-_{2B}$. This signals a {\it non-analytic} harmonic dependence. At the same time, a basic property of the gauge invariant composite operators of the type ${\rm Tr}(\tilde q^{k-n} q^n)$ is their harmonic analyticity (see, e.g., Refs. \cite{EHPSW}), i.e., they must be harmonic polynomials of degree $k$ in $u^+$ only (no $u^-$ are allowed). This is the dynamical expression of the fact that ${\rm Tr}(\tilde q^{k-n} q^n)$ is a highest weight state of an irrep of the R symmetry group SU(2). As we shall see later on, the complete gauge invariant ${\cal N}=2$ amplitude is indeed harmonic analytic. In other words, all the non-analytic terms containing $u^-$ will eventually drop out.  

The second block $J$ has the expression (to lowest order in the $\theta$ expansion)
\begin{equation}\label{defj12}
  J_{102}^{ab} = \frac{4g^2f_{acd}f_{dcb}}{(2\pi)^{6} x_{12}^2} \; [1^-2^-]\rho_1^2\rho_2^2\;.
\end{equation}
Here the blob corresponds to the insertion of the ${\cal N}=2$ SYM Lagrangian ${\cal L} \sim W^2$ which explains the R charge $+4$. As before, the harmonic U(1) charge at points 1 and 2 equals $+1$, and we see the non-analytic factor $[1^-2^-]$.

Note that in the frame (\ref{5}) these building blocks may become simpler or even vanish, for example, $J_{204}\mid_{\theta_{0,2,4}=0} = 0$.

\subsection{The one-loop calculation}\label{pure}

\begin{figure}[tbp]
\begin{center}
\input{diag2.pstex_t}
\end{center}
\caption{Weight $k$}
\label{chargek}\end{figure}
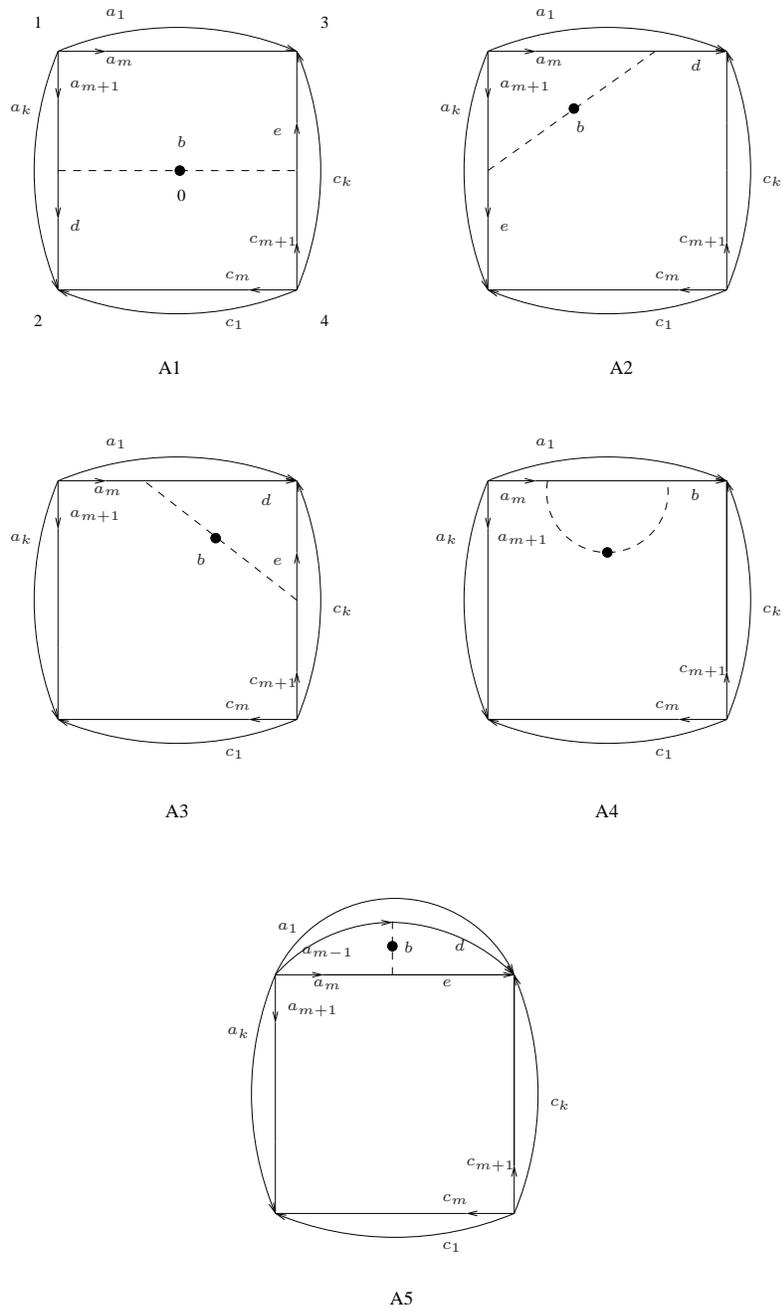

As mentioned earlier, the insertion procedure reduces the one-loop calculation of the four-point correlator $\langle  \tilde q^k q^k q^k \tilde q^k  \rangle$ to a tree level calculation of $\langle {\cal L}\tilde q^k q^k q^k \tilde q^k  \rangle$ (although in the present paper we are mainly interested in the case $k=4$, we keep $k$ arbitrary throughout this subsection). In the frame (\ref{5}) there are only five non-vanishing graphs shown in Fig. \ref{chargek}. All other possible configurations either contain a vanishing building block or a vanishing product of two blocks.
The graphs are labeled by the number $m$ of HM lines connecting, e.g., points 1 and 3. Without loss of generality we can restrict $m$ to run from 1 to $k-1$. Indeed, if $m=0$ the only possible graph A1 becomes one-particle reducible (in fact, it vanishes for colour reasons). If $m=k$ the relevant graphs A4 and A5 become disconnected and correspond to the one-loop correction to the two-point function of the {\it protected} operator ${\rm Tr} (q^k)$, so their sum vanishes. Note also that graph A5 can only exist if $m\geq2$.
 
It is very easy to put together the HM propagators and the two building blocks and to write down the complete five-point amplitude (up to an overall factor): 
\begin{eqnarray}
&&\hskip-1cm \langle {\cal L} \tilde q^k q^k q^k \tilde q^k \rangle|_{\theta_{0,2,4}=0}  \propto \frac{[42]\rho^2_1\rho^2_3}{x^2_{12}x^2_{13}x^2_{24}x^2_{43}} \sum^{k-1}_{m=1} 
\left[ \frac{[12][43]}{x^2_{12}x^2_{43}}\right]^{k-m-1} 
\left[ \frac{[13][42]}{x^2_{13}x^2_{42}}\right]^{m-1}   \nonumber\\
  &\times& \Bigl\{ ({\cal C}^k_m)_{A1}\ [13][21^-][43^-] +  ({\cal C}^k_m)_{A2}\ [13^-][21^-][43] + ({\cal C}^k_m)_{A3}\ [31^-][12][43^-]   \nonumber\\
&&   {} + ({\cal C}^k_m)_{A4}\ [1^-3^-][12][43] -2 ({\cal C}^k_m)_{A5}\ [1^-3^-][12][43] \Bigr\}\;,\label{11}
\end{eqnarray}
where ${\cal C}^k_m$ are the colour factors of the individual graphs.
Each term within the braces in eq. (\ref{11}) is harmonic non-analytic (depends on $1^-, 3^-$). However, the sum of all terms must be analytic. Indeed, according to the discussion in Section \ref{n2inspr}, in the frame (\ref{5}) we expect to find the nilpotent covariant $\Theta^{2222} = [42]^2\rho^2_1\rho^2_3$. The necessary and sufficient condition for this is that the colour factors ${\cal C}$ satisfy the relations (see Appendix \ref{-A})
\begin{equation}\label{colrel}
  ({\cal C}^k_m)_{A2} = ({\cal C}^k_m)_{A3} = ({\cal C}^k_m)_{A4} -2 ({\cal C}^k_m)_{A5} = -({\cal C}^k_m)_{A1} \equiv -c^k_m\;.
\end{equation}
Then we can apply the harmonic identity 
\begin{equation}\label{harid}
  -[13][21^-][43^-] + [13^-] [21^-][43] + [31^-][12][43^-] + [1^-3^-][12][43] = [42]\;,
\end{equation}
which easily follows from the cyclic identity (\ref{harcyc}) and the defining property (\ref{su2harm}) of the harmonics.

In practice, we do not even need to prove harmonic analyticity (which is guaranteed by the superconformal kinematics). Instead, we can profit from it to further simplify our calculation. Indeed, knowing that the five terms within the braces in eq. (\ref{11}) must sum up to the harmonic factor $[42]$, we can identify the harmonic variables $1\equiv 2$ and $3\equiv 4$ within the braces in (\ref{11})
 (but not in the prefactor). Then only graph A1 contributes while all others vanish since, 
e.g., $[11] = 0$. Thus, we obtain the five-point correlator at tree level:
\begin{eqnarray}
&&\hskip-1cm \langle {\cal L} \tilde q^k q^k q^k \tilde q^k \rangle^{\scriptsize\mbox{tree level}} \propto    \nonumber \\ 
&& \frac{[42]^2\rho^2_1 \rho^2_3}{x^2_{12}x^2_{13}x^2_{24}x^2_{43}}\; \sum^{k-1}_{m=1} 
\left[ \frac{[12][43]}{x^2_{12}x^2_{43}}\right]^{k-m-1} 
\left[ \frac{[13][42]}{x^2_{13}x^2_{42}}\right]^{m-1}\; c^k_m \label{8}\\ 
   &\Rightarrow&  \frac{\Theta^{2222}}{x^2_{12}x^2_{13}x^2_{24}x^2_{43}}\; \sum^{k-1}_{m=1} 
\left[ \frac{[12][43]}{x^2_{12}x^2_{43}}\right]^{k-m-1} 
\left[ \frac{[13][42]}{x^2_{13}x^2_{42}}\right]^{m-1}\; c^k_m      \nonumber \\
  &\Rightarrow& \frac{1}{x^2_{12}x^2_{13}x^2_{24}x^2_{43}}\; \frac{\theta^4_0}{x_{01}^2x_{02}^2x_{03}^2x_{04}^2}\; \Bigl[ [13]^2[42]^2 x_{12}^2x_{43}^2 +[12]^2[43]^2x_{13}^2x_{42}^2 
\label{9}\\&&
+ [12][43][13][42]
(x_{14}^2x_{32}^2-x_{13}^2x_{42}^2-x_{12}^2x_{43}^2)\Bigr]
\; \sum^{k-1}_{m=1} 
\left[ \frac{[12][43]}{x^2_{12}x^2_{43}}\right]^{k-m-1} 
\left[ \frac{[13][42]}{x^2_{13}x^2_{42}}\right]^{m-1}\; c^k_m\;. \nonumber  
\end{eqnarray}
In eq. (\ref{8}) we recognize the nilpotent invariant $\Theta^{2222}$ in the frame (\ref{5}). Afterwards, in eq. (\ref{9}) we have switched over to the alternative frame (\ref{xichiral}).  

It remains to carry out the integration over the insertion point according to eq. (\ref{ins}). The Grassmann integral is trivial, $\int d^4\theta_0\; \theta^4_0 = 1$ and that over $x_0$ produces the one-loop scalar box:
\begin{equation}\label{box}
  \int \frac{d^4x_0}{x_{01}^2x_{02}^2x_{03}^2x_{04}^2} = - \frac{i\pi^2}{x^2_{13}x^2_{24}} \Phi(s,t)\;. 
\end{equation}    
Substituting all this in (\ref{ins}) we obtain the final result
\begin{eqnarray}
 &&\hskip-1cm  \langle \tilde q^k q^k q^k \tilde q^k \rangle|^{\scriptsize\mbox{one-loop}} \propto    \nonumber \\  
 && \hskip-0.5cm \Phi(s,t)\; \left\{  \left[ \frac{[12][43]}{x^2_{12}x^2_{43}}\right]^{k} c^k_1 s  +   \left[ \frac{[12][43]}{x^2_{12}x^2_{43}}\right]^{k-1}\left[ \frac{[13][42]}{x^2_{13}x^2_{42}}\right]  \left(c^k_2 s + c^k_1 (t-s-1)\right) \right. \nonumber\\
  && +\sum^{k-2}_{m=2}    \left[ \frac{[12][43]}{x^2_{12}x^2_{43}}\right]^{k-m} \left[ \frac{[13][42]}{x^2_{13}x^2_{42}}\right]^{m} \left(c^k_{m+1} s + c^k_m (t-s-1) + c^k_{m-1}  \right) \nonumber\\
  && \left. {}+   \left[ \frac{[12][43]}{x^2_{12}x^2_{43}}\right]\left[ \frac{[13][42]}{x^2_{13}x^2_{42}}\right]^{k-1}  \left( c^k_{k-1} (t-s-1) + c^k_{k-2}  \right)  + \left[ \frac{[13][42]}{x^2_{13}x^2_{42}} \right]^{k} c^k_{k-1}   \right\}  \;. \label{1loopgen}
\end{eqnarray}
Note that eq. (\ref{1loopgen}) reproduces the well-known result for $k=2$ \cite{GPS,EHSSW,BKRS2}. 

The colour factors $c^k_m$ are not difficult to compute in the large $N$ limit  (see Appendix \ref{-A}):
\begin{equation}\label{colorn}
  c^k_m|_{N \to \infty} \approx -2 k^4 N^{2k-1}\; ,
\end{equation}
{i.e.},  they are independent of $m$.
We restrict our further analysis to the most interesting case $k=4$ and just remark that the case $k=3$ is treated in a similar manner. 

Inserting the colour factors into eq. (\ref{1loopgen}),  setting $k=4$ and comparing with (\ref{1'}) we derive the values of 5 of the 15 coefficients in the ${\cal N}=4$ amplitude (\ref{Ampl}) (up to an overall factor):
\begin{equation}\label{values}
  a_1 \sim N^7s\Phi\;, \quad a_2 \sim N^7\Phi\;, \quad b_1 \sim N^7(t-1)\Phi\;, \quad b_4 \sim N^7(t-s)\Phi\;, \quad c_1 \sim N^7t\Phi\;.
\end{equation}
Next we restore the overall normalization of the four-point amplitude corresponding to operators with canonically normalized two-point functions. Finally, comparing the properly normalized one-loop result (\ref{values}) to the general form (\ref{a})-(\ref{c}), we see that {\it the independent functions parametrizing the one-loop amplitude coincide}:
\begin{equation}\label{result}
  \qquad {\cal F}(s,t) = {\cal G}(s,t)=-\frac{4^2}{8\pi^2}\frac{\lambda}{ N^2}\Phi(s,t) \;.
\end{equation}

We point out that the above effect is only true in the large $N$ limit. Indeed, the exact expressions for the colour coefficients 
differ at finite $N$.\footnote{We thank Silvia Penati and Alberto Santambrogio for a discussion on this point.} On the other hand, in our one-loop calculation we have only dealt with single-trace operators ${\rm Tr}{\cal W}^4$. In fact, $k=4$ is the first case where a $\half$-BPS operator can be realized as a single- as well as a double-trace operator (for $k=2,3$ there can only be single traces if the gauge group is SU($N$)). It is plausible that by choosing the appropriate mixing coefficient one can always achieve that the two functions coincide, as in (\ref{result}).

Another remark concerns the U(1) projection (\ref{8''d}). Using (\ref{values})  we find
\begin{equation}\label{8'''}
  \langle \bar W^4| W^4|W^4|\bar W^4 \rangle|^{\scriptsize\mbox{one-loop}}_{N \to \infty} \propto  \frac{\Phi(s,t)}{x^8_{13}x^8_{42}} \frac{t}{s}\left(1 + \frac{1}{s} + \frac{1}{s^2} \right)\; . \end{equation}
Recently, the large $N$ one-loop amplitude corresponding to the U(1) projection of the
four-point amplitude of $\half$-BPS operators of arbitrary weight $k$
was computed in Ref. \cite{BKPSS}  in the context of the so-called pp-wave limit in ${\cal N}=4$ SYM. It is easy to see that the formula (\ref{8'''}) agrees
with the results of Ref. \cite{BKPSS}. We stress, however, that according 
to eq. (\ref{8''d}) the knowledge of this U(1) projection alone is not enough to 
identify the individual functions $\cF$ and $\cG$.

\section{Four-point amplitude from Type IIB supergravity}\label{gravity}
To obtain the four-point amplitude from supergravity we first have 
to identify the field theory operator with the corresponding 
supergravity field. As mentioned above, 
the $\half$-BPS multiplets of weight 4 
appear in gauge theory as linear combinations 
(mixtures) of a single- and a double-trace operators. We recall that the operator mixing possible 
on the gauge theory side corresponds to (non-linear) redefinitions 
of the fields in the supergravity Lagrangian. 
It is however known \cite{AF4} that for a four-point amplitude 
of the {\it regular type} (i.e. neither extremal nor subextremal), 
which is our case here,
the effect of the operator mixing is suppressed in the large 
$N$ limit. Thus, we can safely identify the single-trace operator ${\cal O}^I$
with the KK mode $s^I$ (with vanishing AdS mass) 
arising in the spectrum of the compactified Type IIB supergravity 
and use the effective action of Ref. \cite{AF2} to derive the corresponding 
four-point amplitude.

The KK spectrum of the compactified IIB supergravity was found in Refs.  \cite{KRN}. The scalar $s_k$, $k\geq 2$, transforming in the irrep $[0,k,0]$
is the superconformal primary of a short supermultiplet with top spin 2:
\bea
[s,A_{\mu},\varphi_{\mu\nu},\phi,C_{\mu},t,...]_k \, .
\eea    
The dots stand for a number of other fields which do not appear in the cubic couplings together with two fields $s_k$.
All the fields involved in the cubic couplings are bosonic and have vanishing $U(1)_Y$ charge. In Table 1 we list their SU(4) representation labels, AdS mass $m^2$ and the conformal dimension $\Delta$ of the corresponding operator in the dual CFT.
\vskip 0.7cm
\begin{center}
\begin{tabular}{|c|c|c|c|c|c|c|}
\hline
Field & $s_k$ & $A_{\mu,k}$ & $C_{\mu,k}$ & $\phi_k$ & $t_k$ & $\phi_{\mu\nu~k}$ \\
\hline
Irrep & $[0,k,0]$ & $[1,k-2,1]$ & $[1,k-4,1]$ & $[2,k-4,2]$  & $[0,k-4,0]$ & $[0,k-2,0]$ \\
\hline
$m^2$ & $k(k-4)$ &  $k(k-2)$    & $k(k+2)$    & $k^2-4$      & $k(k+4)$ & $k^2-4$  \\
\hline
$\D $ & $k$      &  $k+1$       & $k+3$       & $k+2$        & $k+4$    & $k+2$   \\
\hline
\end{tabular}
\end{center}
\begin{center}
\begin{tabular}{l}
Table 1. Components of the $\half$-BPS multiplet 
mediating the interactions 
of two \\ superconformal primaries $s_k$. 
\end{tabular}
\end{center}
It is worthwhile noting that the generic $\half$-BPS multiplets 
start with $k=4$. The case $k=2$ corresponds to the ``ultrashort" graviton multiplet whose SYM dual is the stress-tensor multiplet. The multiplet with $k=3$ is also shorter than that for $k\geq 4$. Indeed, the scalars $\phi$ and $t$, as well as the vector $C_{\mu}$ appear only starting with $k=4$.

Before starting the computation, we first discuss the relevant part of the effective 
five-dimensional action obtained by compactifying IIB supergravity 
on an $AdS_5\times S^5$ background \cite{LMRS}, \cite{AF2}: 
 \bea S=\frac{N^2}{8\pi^2}\int d^5x
\sqrt{g_a} \left( {\cal L}_2+{\cal L}_3+{\cal L}_4 \right) \, , \eea which is a sum of quadratic, cubic and quartic terms. Here $g_a$ is the
determinant of the Euclidean AdS metric
$ds^2=\frac{1}{z_0^2}(dz_0^2+dx^a dx^a)$, $a=1,2,3,4$.

The quadratic Lagrangian reads as follows:
\bea
\nonumber
{\cal L}_2&=&\sum_{k=2,4,6}N_k^s\left(\nabla_{\mu} s_{ k}\nabla^{\mu} s_{ k}
+m^2_s(k) s_{ k}^2  \right)+
\frac{3}{2}\left(\nabla_{\mu} \phi\nabla^{\mu} \phi
+m^2_{\phi}\phi^2  \right)\\
\nonumber
&+&\sum_{k=2,4,6}
\Big[\frac{1}{2}F_{\mu\nu,{ k}}^2(A)+m^2_A(k)A_{\mu,{ k}}^2\Big]
+\left(F_{\mu\nu}^2(C)+m^2_CC_{\mu }^2\right) \\
\nonumber
&+&\sum_{k=2,4,6}N_k^{\varphi}\Big[
\frac{1}{4}\nabla_{\rho}\varphi_{\mu\nu~k}\nabla^{\rho}\varphi^{\mu\nu}_k
-\frac{1}{2}\nabla_{\mu}\varphi_{\mu\rho~k}\nabla^{\nu}\varphi_{\nu\rho~k}
+\frac{1}{2}\nabla_{\mu}\varphi_{\rho~k}^{\rho}\nabla_{\nu}\varphi^{\mu\nu}_k \\
\label{quad}
&-&\frac{1}{4}\nabla_{\rho}\varphi_{\mu~k}^{\mu}\nabla^{\rho}
\varphi^{\nu}_{\nu~k}
+\frac{1}{4}(k^2-6)\varphi_{\mu\nu~k}\varphi^{\mu\nu}_k
-\frac{1}{4}(k^2-2)(\varphi_{\mu~k}^{\nu})^2
\Big] \, .
\eea
For the sake of clarity we have suppressed the summation over the representation index of the corresponding irreps. The normalization constants $N_k^s$ and $N_k^{\varphi}$ are chosen to be 
\bea
N^s_2=\frac{1}{4}\, ,~~~N_4^s=\frac{3}{2}\, , ~~~N^s_6=15 \, , ~~~
\nonumber
N_2^{\varphi}=1\, , ~~~N_4^{\varphi}=6  \, ,~~~N_6^{\varphi}=\frac{5}{3}\, .
\eea

In eq. (\ref{quad}) we present only the fields which participate in the 
cubic coupling with two scalars $s_4$. The masses and the SU(4) irreps of these fields are read off from Table 1,  in particular, $m^2_{\phi}=12$ and $m^2_C=24$.
In total, the Lagrangian involves eleven fields which are organized into three 
$\half$-BPS families labeled by $k=2,4,6$.

To write down the cubic Lagrangian we introduce the Clebsh-Gordon 
coefficients 
$$\langle C^1C^2C^3_{[a_1,a_2,a_3]} \rangle$$
for the tensor product of two irreps $[0,4,0]$ (the first two indices) and one 
$[a_1,a_2,a_3]$ (the third index). For the definitions and normalizations of the 
$C$-tensors involved, see \cite{AF2,ADOS}.
According to Ref. \cite{AF2}, we  have\footnote{In comparison to \cite{AF2} we made suitable
rescalings of the fields to fit the normalization of the quadratic action. To simplify the notation we
also identify $s\equiv s_4$.} 
\bea
\nonumber
&&{\cal L}_3=-36\langle C^1C^2C^3_{[0,2,0]} \rangle s^1s^2s^3_2
-12\langle C^1C^2C^3_{[1,0,1]}\rangle s^1\nabla^{\mu}s^2 A_{\mu,2}
-\frac{3}{2}\d^{12}T^{12~\mu\nu}_{2} \varphi_{\mu\nu , 2} \\
\nonumber
&&-96\langle C^1C^2C^3_{[0,4,0]} \rangle s^1s^2s^3
-36\langle C^1C^2C^3_{[1,2,1]}\rangle s^1\nabla^{\mu}s^2 A_{\mu,4}
-\frac{36}{5}\langle C^1C^2C^3_{[0,2,0]} \rangle T^{12~\mu\nu}_{4}\varphi_{\mu\nu , 4}^3
\\
\nonumber
&&-18\langle C^1C^2C^3_{[2,0,2]} \rangle s^1s^2\phi^3
-\frac{12}{5}\langle C^1C^2C^3_{[1,0,1]}\rangle s^1\nabla^{\mu}s^2 C_{\mu} 
\\
\nonumber
&&
-1080 \langle C^1C^2C^3_{[0,6,0]} \rangle s^1s^2s^3_6
-36 \langle C^1C^2C^3_{[1,4,1]}\rangle s^1\nabla^{\mu}s^2 A_{\mu,6}
-3\langle C^1C^2C^3_{[0,4,0]} \rangle T^{12~\mu\nu}_{6} \varphi_{\mu\nu , 6}^3.
\eea
Here we use the notation:
\bea
\nonumber
T^{12}_{\mu\nu~k}=\frac{1}{2}\nabla_{\mu}s^1\nabla_{\nu}s^2+
\frac{1}{2}\nabla_{\nu}s^1\nabla_{\mu}s^2-
\frac{g_{\mu\nu}}{2}\Big(\nabla_{\rho} s^1 \nabla^{\rho}s^2-\frac{1}{2}(k^2-4)s^1s^2  \Big)\, .
\eea
The first line of the expression for ${\cal L}_3$ shows the contribution of the fields from the $k=2$ multiplet. Analogously, the second and the third lines give the contribution of the $k=4$ multiplet, while the fourth line corresponds to $k=6$.
We thus see that the cubic Lagrangian (and therefore (\ref{quad})) involves all the descendents from the $k=2$ and $k=4$ multiplets which are allowed by the tensor product decomposition (\ref{tp}) and by the $U(1)_Y$ selection rule, except for the scalar $t$. This is due to the fact that the conformal dimension of the   operator dual to $t$ is 8, i.e. the corresponding cubic coupling is extremal and therefore it vanishes.


\begin{picture}(30000,15000)
\drawline\scalar[\W\REG](25000,5000)[5]
\global\advance\pmidx by -3300
\global\advance\pmidy by 500
\put(\pmidx,\pmidy){$[s,A_{\mu},\varphi_{\mu\nu},\ldots]_k$}
\drawline\fermion[\SW\REG](\particlebackx,\particlebacky)[4500]
\global\advance\pbackx by -1200
\put(\pbackx,\pbacky){$s_4$}
\drawline\fermion[\NW\REG](\scalarbackx,\scalarbacky)[4500]
\global\advance\pbackx by -1200
\put(\pbackx,\pbacky){$s_4$}
\drawline\fermion[\SE\REG](\scalarfrontx,\scalarfronty)[4500]
\global\advance\pbackx by 300
\put(\pbackx,\pbacky){$s_4$}
\drawline\fermion[\NE\REG](\scalarfrontx,\scalarfronty)[4500]
\global\advance\pbackx by 300
\put(\pbackx,\pbacky){$s_4$}
\end{picture}
\begin{center}
\begin{tabular}{l}
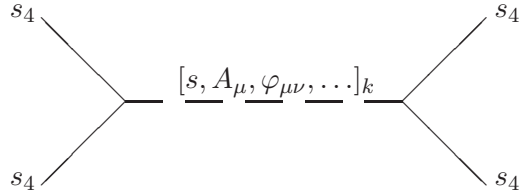
 Figure 5: Exchange graphs contributing to the four-point function of operators ${\cal O}^I$.
For \\ every $k=2,4,6$ the fields exchanged belong to the 
same multiplet $[s,A_{\mu},\varphi_{\mu\nu},\ldots]_k$. 
\end{tabular}
\end{center}

\vskip 0.5cm
Not all of the descendents of the $k=6$ multiplet appear in ${\cal L}_3$, 
either because of incompatibility with the selection rule implied by (\ref{tp}) 
or because the corresponding  coupling is extremal. Of course, all these missing descendent fields will enter the cubic Lagrangian involving the fields $s_k$ with $k>4$.

Finally, we need to extract the relevant contact interactions from the general quartic action of Ref. \cite{AF2}. To this end (as well as to compute the contribution of the exchange graphs to the four-point amplitude (\ref{Ampl})) it is useful to re-expand the elements of the so-called ``OPE basis'' given by 
\bea
\label{OPEbasis}
\langle C^1C^2C^5_{[a_1,a_2,a_3]} \rangle\langle C^3C^4C^5_{[a_1,a_2,a_3]} \rangle  
\eea  
(the summation over the representation index at the fifth leg is assumed), over the 15 elements of the ``propagator basis'' (\ref{propbasis}) (see Figure 1). 
Here at the legs 1, 2, 3 and 4 we have the external irrep $[0,4,0]$;
in general, for an operator of weight $k$ it would be $[0,k,0]$. 

The problem of re-expanding the OPE basis over the propagator basis 
is solved in Appendix \ref{A}. Then following the same steps as in \cite{ADOS}, that is, expressing the quartic effective action of Ref. \cite{AF2} in the propagator basis and integrating by parts, we arrive at the very simple expression 
\bea
{\cal L}_4&=&-
\frac{9}{5}\left(C^{1423}+3\Upsilon^{1234}-4\Omega^{1234}\right)s^1\nabla_{\mu} s^2 s^3 \nabla^{\mu} s^4\\
\nonumber
&-&\frac{9}{5}\left(8C^{1234}-234\Upsilon^{1234}-62\Omega^{1234}+3\d^{12}\d^{34}\right)
s^1s^2s^3s^4 \, .
\eea
Again, as in the case of the weight 3 operators, the quartic terms
with four derivatives disappear completely, i.e. {\it the final action appears to be of the sigma model type}. This is another explicit example supporting the conjecture that the extension of the five-dimensional gauged ${\cal N}=8$ supergravity by inclusion of the KK modes of the compactified IIB theory can be described by some sigma model. However, to answer this question definitely one has to analyze the 
effective action corresponding to the CFT operators of unequal charges.
We hope to return to this interesting issue in future.  

With the relevant interacting Lagrangian at hand one can now compute the corresponding four-point amplitude. Since the calculations of this kind are well-described in the literature \cite{DHFMMR,DHMMR,AF3}, we will not repeat them here.\footnote{The scalar, vector and the graviton exchange graphs are treated by the general technique of Ref. \cite{dHFR}. The contribution of the exchange graphs of massive symmetric tensors
was established in Appendix E of Ref. \cite{ADOS}.} We present the coefficient functions of the resulting four-point amplitude in Appendix \ref{B}. Not surprisingly, in comparison with the cases of operators of weight 2 and 3, our present amplitude turns out more involved due to the larger number of various exchange graphs contributing.  

Having found the supergravity induced four-point amplitude for the $\half$-BPS operators of weight 4, we can now verify whether it has the structure (\ref{a})-(\ref{d}) predicted by the partial non-renormalization theorem of Section \ref{general}, i.e. that all the coefficient functions are expressed in terms of two and only two {\it a priori} independent functions ${\cal F}(s,t)$ and ${\cal G}(s,t)$ (\ref{cr}). As discussed in Ref. \cite{ADOS}, there are two different ways to  perform this check. Firstly, one can rewrite all 
the $D$-functions describing the coefficients  $a_i$ and $b_i$ via a single generating function of the conformal cross-ratios, $\Phi(s,t)$. After this has been done, the verification of eqs. (\ref{a})-(\ref{d}) becomes straightforward. We omit the details of the corresponding calculation and just state that our supergravity induced four-point amplitude has indeed exactly the form predicted by the partial non-renormalization theorem. The second but equivalent way is to rewrite the coefficient functions via the $\oD$-functions of Ref. \cite{DO} (which are a variant 
of the $D$-functions of Ref. \cite{DHFMMR}) and to exploit the  various identities among them to show how partial non-renormalization works. This  approach seems to be the most efficient one, as it also allows us to achieve a dramatic simplification of the original coefficient functions. Proceeding in this way, we have not only been able to identify the basic functions $\cF$ and $\cG$ describing our four-point amplitude, but also to write them down in a form which makes their crossing symmetry properties explicit:\footnote{See Appendix \ref{B}, where some steps of this derivation are outlined.}
\bea
\label{simpleFG}
\cF(s,t)&=&-\frac{4}{N^2}\Big[2{\oD}_{2246}+2s{\oD}_{3346}+s^2{\oD}_{4446}\Big] \, ,\\
\nonumber
\cG(s,t)&=&-\frac{16}{N^2}s\Big[
\oD_{4222}+(\oD_{4233}+\oD_{4323})+(\oD_{4244}+\oD_{4424}+5\oD_{4334})-\oD_{5335}+\oD_{4446}
\Big] \, .
\eea
Indeed, the symmetry relations  (\ref{cr}) readily follow from the ones for the $\oD$-functions  
\bea
\label{trans}
\oD_{\D_1\D_2\D_3\D_4}(s/t,1/t)&=&t^{\Sigma-\D_4}\oD_{\D_2\D_1\D_3\D_4}(s,t) \, , \\
\nonumber
\oD_{\D_1\D_2\D_3\D_4}(1/s,t/s)&=&s^{\Sigma-\D_4}\oD_{\D_1\D_3\D_2\D_4}(s,t) \, , 
\eea
where $\Sigma=1/2\sum_i\D_i$.   
From the representation  (\ref{simpleFG}) one can also conclude that the 
functions  ${\cal F}$ and ${\cal G}$ are {\it distinctly  different} as there is no way to reduce the one to the other by using identities between $\oD$-functions with different indices. To see this we recall that these algebraic identities relate the $\oD$-functions with the same value of $\Sigma$ (up to $\oD$-functions with lower $\Sigma$). The top component of $\cF$ with $\Sigma=9$ is proportional to $s^2\oD_{4446}$, while the top component of $\cG$ with the same $\Sigma$ gives $s\oD_{4446}$, i.e. they differ by a factor of $s$. 

Finally, we note that in order to obtain an agreement with the partial non-renormalization theorem, we have to assume that our supergravity induced four-point amplitude has a ``free''
part where the coefficients $a_i,b_i,c_i,d_i$ are precisely those from eqs.  (\ref{free}). 
\section{Summary and conclusions}
In this section we collect and discuss the known results about the four-point amplitudes of $\half$-BPS operators of different weights, both in the perturbative and in the supergravity regimes.  

The coefficient functions for $k=2,3,4$ have the following form (we give one representative of each crossing-equivalence class, e.g., $a_1,b_1,c_1,d_1$ for $k=4$, cf. eq. (\ref{Ampl}) and Figure 1):  \pagebreak
\begin{itemize}
\item{Weight 2} 
\bea
\label{2sugra}
a(s,t) &=& 1 + s{\cal F}_2(s,t) \nonumber\\
b(s,t) &=& \frac{4}{N^2} + (s-t-1){\cal F}_2(s,t)
\eea
\item{Weight 3}
\bea
\label{3sugra}
a(s,t) &=& 1 + s{\cal F}_3(s,t) \nonumber\\
b(s,t) &=& \frac{9}{N^2} + (t-s-1){\cal F}_3(s,t)  +   \frac{s}{t}{\cal F}_3({1}/{t},{s}/{t})\\
c(s,t) &=& \frac{18}{N^2} + (s-t-1){\cal F}_3(s,t)  + (t-s-1){\cal F}_3(t,s) + \frac{1-s-t}{t}{\cal F}_3({1}/{t},{s}/{t}) \nonumber\eea
\item{Weight 4} \\ 
\bea
\label{4sugra}
a(s,t) &=& 1 + s{\cal F}_4(s,t) \nonumber\\
b(s,t) &=& \frac{16}{N^2} + (t-s-1){\cal F}_4(s,t) + s\cG_4(s,t)  \nonumber\\
c(s,t) &=& \frac{16}{N^2} + {\cal F}_4(s,t) + \frac{s}{t}{\cal F}_4({1}/{t},{s}/{t})  + (t-s-1){\cal G}_4(s,t) \\
d(s,t) &=& \frac{32}{N^2} + (s-t-1){\cal F}_4(s,t)  \nonumber\\
&& {} + s {\cal G}_4(t,s) + (1-t-s){\cal G}_4(s,t) + \frac{t-s-1}{t}{\cal G}_4({s}/{t},{1}/{t}) \nonumber
\eea
\end{itemize}
Here the constant terms coincide with the coefficients in the free amplitude calculated in the large $N$ limit for canonically normalized single-trace operators. The functions ${\cal F}_k$ ($k=2,3,4$) and ${\cal G}_4$ have the symmetry properties
\bea
\label{crsum}
{\cal F}(s,t)=1/t\,{\cal F}(s/t,1/t) \, , ~~~~~{\cal G}(s,t)=1/s\,{\cal G}(1/s,t/s)\, .
\eea   
In addition, the function ${\cal F}_2$ has the extra symmetry
\begin{equation}\label{extrasym}
  {\cal F}_2(s,t) = {\cal F}_2(t,s)\, .
\end{equation}

Now we list the explicit expressions for ${\cal F}$ and ${\cal G}$, both in the perturbative and supergravity regimes. At one loop and in the large $N$ limit we have \bea
\cF^{\rm 1-loop}_k(s,t) &=& -\frac{k^2}{8\pi^2}\frac{\lambda}{N^2}\Phi^{(1)}(s,t) \, , \label{1loopres}\\
 {\cal G}^{\rm 1-loop}_4(s,t) &=& {\cal F}^{\rm 1-loop}_4(s,t)\, , \label{surprise}
\eea 
where $\Phi^{(1)}(s,t)\equiv \Phi(s,t)$ is the one-loop scalar box integral.

The fact that for  $k=2,3$ the amplitude is determined by a single function,  ${\cal F}_k$, follows from the partial non-renormalization theorem (the insertion procedure). However, for $k=4$ our general field-theoretic arguments can only predict the appearance of two conformal invariant functions, $\cF_4$ and $\cG_4$. The result  $\cF^{\rm 1-loop}_4=\cG^{\rm 1-loop}_4$ is of genuine {\it dynamical origin}. The universal dependence on $k$ in eq. (\ref{1loopres}) allows us to believe that even for arbitrary $k$ the one-loop four-point amplitude is described in terms of a single function. We also point out that already for $k=3$ we observe a certain degeneracy of the one-loop amplitude (which holds at finite $N$ as well), namely, the function ${\cal F}_3$ from (\ref{1loopres}) has the extra symmetry ${\cal F}_3(s,t) = {\cal F}_3(t,s)$, compared to the general requirement (\ref{crsum}).

In the supergravity regime we find
\begin{eqnarray}
  {\cal F}^{\rm SG}_2(s,t)&=& -\frac{4}{N^2} \oD_{2224}\label{2sugraSG}\\
  {\cal F}^{\rm SG}_3(s,t)&=& -\frac{9}{N^2}\left(\oD_{2235}+s\oD_{3335}\right) \label{3sugraSG}\\
  {\cal F}^{\rm SG}_4(s,t)&=& -\frac{4}{N^2}\left(2\oD_{2246}+2s\oD_{3346}+s^2\oD_{4446}\right)\label{4sugraSG}\\
  {\cal G}^{\rm SG}_4(s,t)&=& -\frac{16}{N^2}s\Big[
\oD_{4222}+(\oD_{4233}+\oD_{4323})   \nonumber\\ 
    &&\qquad      +(\oD_{4244}+\oD_{4424}+5\oD_{4334})-\oD_{5335}+\oD_{4446}
\Big]\, .\label{4sugra'}
\end{eqnarray}
The coefficients (\ref{2sugraSG}) and (\ref{3sugraSG}) were found in Refs. \cite{AF3,EPSS,DO} and \cite{ADOS}, respectively. The expressions (\ref{4sugraSG}) and (\ref{4sugra'}) constitute a new result. Using the symmetry properties (\ref{trans}) we can readily see that for $k=2,3,4$, the function ${\cal F}_k^{\rm SG}$ obeys the relation (\ref{crsum}), and for $k=2$ the function ${\cal F}_2^{\rm SG}$ is in addition symmetric, as required by (\ref{extrasym}).

These examples suffice to illustrate the essential difference between the perturbative and the supergravity results.  While at one loop increasing the weight results in a simple change of the coefficient $k^2$, in supergravity it leads to an increased number of $\oD$-functions; the $\oD$-function with the highest value of $\Sigma=2k+1$ occurring in $\cF_k$ is $\oD_{k,k,k,k+2}$. In this context it is amusing to note that if there would exist a gauge-invariant $\half$-BPS operator of weight $k=1$ (corresponding to the singleton multiplet of the AdS supergravity), then the matching supergravity-induced amplitude could be proportional to $\oD_{1113}$.

In order to better understand the degeneracy 
phenomenon observed here and its implications for the AdS/CFT duality conjecture, 
it is necessary to study the higher-order perturbative corrections. So far,  
the two-loop (order $\lambda^2$) result for the four-point amplitude is available only for the case $k=2$ \cite{ESS,BKRS4}  and it reads
\begin{eqnarray}\label{2loop}
  \cF^{\rm 2-loop}_2(s,t)&=&\frac{2^2 }{4\cdot (2\pi)^4 } \frac{\lambda^2}{N^2}\Biggl[\frac{1}{4}(s+t+1)[\Phi^{(1)}(s,t)]^2 \\
\nonumber
&+&
 \frac{1}{s}\Phi^{(2)}(t/s,1/s)+ \Phi^{(2)}(s,t)+ \frac{1}{t}\Phi^{(2)}(s/t,1/t)
 \Biggr] \, , 
\end{eqnarray}
where we have exhibited the dependence on the weight, $2^2=k^2$, and where $\Phi^{(2)}(s,t)$ is the standard two-loop scalar box integral.
Thus, the next step would be to find the four-point amplitude for $k=3,4$ at two loops \cite{APSS}, in order to check if the one-loop degeneracy (the extra symmetry of $\cF_3$ and the equality $\cF_4=\cG_4$) is lifted at this order or not. 

The universal presence of the free (constant) part in the amplitude coefficients (\ref{2sugra})-(\ref{4sugra}) deserves a special comment. In the perturbative amplitude the coupling $\lambda$ provides a natural splitting into a free and a quantum parts. However, in the supergravity amplitude there is no such parameter. Yet, we can argue that at least part of the free amplitude should appear unchanged in the supergravity regime, if we believe in the AdS/CFT duality. The point is that any amplitude is subject to superconformal Ward identities which imply that some of the coefficient functions actually depend on a {\it single variable} \cite{EPSS,DO,HH}. Further, the insertion procedure tells us that these functions do not get quantum corrections (in other words, their OPE contains only protected multiplets). Then the part of the free amplitude which can be rewritten in terms of such functions should also be present in the supergravity regime. This is indeed the case, and we consider this fact as yet another non-trivial confirmation of the AdS/CFT duality. 

It is worthwhile noting that the free part of the perturbative amplitude has been calculated under the assumption that the $\half$-BPS operators are realized as single traces. At the same time, the case $k=4$ is the first one when mixing between single- and double-traces is possible. According to the discussion above, any such mixing must be suppressed in the large $N$ limit, otherwise we would not find an agreement with the constant part of the supergravity amplitude. We plan to give a detailed discussion of the structure of the free amplitude in a forthcoming paper \cite{APSS}.

The partial non-renormalization theorem predicting two independent functions in the case $k=4$ has a non-perturbative nature. Therefore it must also apply to the correlator computed in an instanton background. For $k=2$, where partial non-renormalization predicts a single function, this was confirmed in Refs. \cite{BGKR,DKMV} where the corresponding function was found to be 
\begin{equation}\label{inst}
  \cF_2^{\rm inst}(s,t)\sim st \oD_{4444} \, .
\end{equation}      
It is very interesting to generalize the instanton computations of Refs. \cite{BGKR,GK} 
to the cases $k=3,4$ both in the gauge theory and by using the low energy superstring
effective action and to see if and how the one-loop degeneracy is lifted by the instanton effects.    

To shed some light on the r\^ole of the functions $\cF_4$ and $\cG_4$ it is useful to study the corresponding operator product expansion (OPE). Just by using Wick contractions it is easy to see that the OPE of two weight $k$ operators ${\cal O}^{(k)}{\cal O}^{(k)}$ must have a heredity property, that is, if a superconformal primary operator participates in the OPE of two $\half$-BPS operators of weight $k-1$, it should also appear in the OPE of two operators of weight $k$. In particular, the OPE ${\cal O}^{(2)}{\cal O}^{(2)}$ contains an infinite tower of twist 2 operators of increasing spin, whose lowest member is the Konishi scalar. So, this tower must be present in the perturbative OPE derived for any other value of $k$. We have analyzed the OPE for $k=4$ under the assumption $\cF_4 = \mu\Phi$ and $\cG_4 = \nu\Phi$ where $\mu,\nu$ are different numerical coefficients (this is a slight deformation of the actual one-loop result). We found that only $\mu$ is fixed by the requirement to reproduce the known one-loop anomalous dimensions for the whole tower of twist 2 fields. Therefore, the function $\cG_4$ seems to be responsible for creating the anomalous dimensions of higher twist fields. One should try to better understand the implications of this observation. Similar test can be carried out at two loops as well \cite{APSS}.
                                       
\newpage
                                             
\section*{Acknowledgements} We are grateful to N. Beisert, F. Dolan, S. Frolov, S. Kovacs, H. Osborn, S. Penati, A. Petkou, J. Plefka, A. Santambrogio,  M. Staudacher and S. Theisen for many useful discussions. G.A. and E.S. wish to thank the LAPTH and the Max Planck Institute f\"ur Gravitationsphysik, respectively, for the warm hospitality. G.A. was supported in part by the European Commission RTN programme HPRN-CT-2000-00131 and by RFBI grant N02-01-00695.

\section*{Appendices}
\appendix
\setcounter{equation}0

\section{Colour factors}\label{-A} 

Here we prove the colour identities (\ref{colrel}) and calculate the colour factors in the large $N$ limit.   

The explicit expressions for the colour factors are as follows:
\begin{eqnarray}
 ({\cal C}^k_m)_{A1} &=& \frac{f_{a_{m+1}bd}f_{c_{m+1}be}}{[(k-m-1)!m!]^2} \nonumber\\
  &&  (a_1\cdots a_k)(c_1\cdots c_k)
 (a_1\cdots a_m e c_{m+2} \cdots c_k)(c_1\cdots c_m d a_{m+2} \cdots a_k)  \nonumber\\
 ({\cal C}^k_m)_{A2} = ({\cal C}^k_m)_{A3} &=& \frac{f_{a_{m}bd}f_{a_{m+1}be}}{(k-m)!(k-m-1)!m!(m-1)!} \nonumber\\
  &\times&  (a_1\cdots a_k)(c_1\cdots c_k)(a_1\cdots a_{m-1} d c_{m+1} \cdots c_k)(c_1\cdots c_m e a_{m+2} \cdots a_k) \nonumber\\
 ({\cal C}^k_m)_{A4} &=&  \frac{f_{a_{m}de}f_{deb}}{[(k-m)!]^2 m!(m-1)!}  \label{13} \\
  &\times&  (a_1\cdots a_k)(c_1\cdots c_k)(a_1\cdots a_{m-1} b c_{m+1} \cdots c_k)(c_1\cdots c_m a_{m+1} \cdots a_k)  \nonumber\\
 ({\cal C}^k_m)_{A5} &=& \frac{f_{a_{m-1}bd}f_{a_{m}be}}{[(k-m)!]^2 m!(m-2)!2!}  \nonumber\\
  &\times&  (a_1\cdots a_k)(c_1\cdots c_k)(a_1\cdots a_{m-2} d e c_{m+1} \cdots c_k)(c_1\cdots c_m a_{m+1} \cdots a_k)\nonumber  
\end{eqnarray} 
where 
\begin{equation}\label{12}
  (a_1\cdots a_k) \equiv {\rm Tr}(t_{(a_1} \cdots t_{a_k)})
\end{equation}
is the symmetrized (without $1/k!$) trace of $k$ generators of the colour group. The combinatorial coefficients are needed to avoid overcounting identical HM lines. 

Now, let us first show that $({\cal C}^k_m)_{A1} = -({\cal C}^k_m)_{A2}$. To this end we first replace the structure constants by commutators under the traces, according to the Lie algebra $[t_a,t_b]=if_{abc}t_c$:
\begin{eqnarray}
({\cal C}^k_m)_{A1} &=& -\frac{1}{[(k-m-1)!m!]^2}(a_1\cdots a_k)(c_1\cdots c_k) \nonumber\\
  &\times& (a_1\cdots a_m [b,c_{m+1}] c_{m+2} \cdots c_k)(c_1\cdots c_m [b,a_{m+1}] a_{m+2} \cdots a_k) \label{131}\\
 ({\cal C}^k_m)_{A2} &=& - \frac{1}{{(k-m)!(k-m-1)!m!(m-1)!}}(a_1\cdots a_k)(c_1\cdots c_k) \nonumber\\                            
 &\times&  (a_1\cdots a_{m-1} [b,a_m] c_{m+1} \cdots c_k)  (c_1\cdots c_m [b,a_{m+1}] a_{m+2} \cdots a_k)  \nonumber 
\end{eqnarray}  
(the symmetrization does not involve the commutator). Next, following \cite{dHFS} we use the identity
\begin{equation}\label{skiba}
  \sum_{p=1}^q {\rm Tr}\left(M_1 \cdots [N,M_p] \cdots M_q  \right) = 0
\end{equation}
to write down
\begin{eqnarray}
 &&\hskip-1cm - (a_1\cdots a_m [b,c_{m+1}] c_{m+2} \cdots c_k) =  \label{sk2} \\
 &&  (a_1\cdots a_m c_{m+1}[b,c_{m+2}] \cdots c_k) + \cdots + (a_1\cdots a_m c_{m+1} c_{m+2} \cdots [b,c_k]) \nonumber\\
  && {}+ ([b,a_1]\cdots a_m c_{m+1} c_{m+2} \cdots c_k) + \cdots + (a_1\cdots [b,a_m] c_{m+1} c_{m+2} \cdots c_k) \;.  \nonumber
\end{eqnarray}
Recalling the total symmetrization over all the indices $a_p$ and $c_p$ from (\ref{131}), we see that all the $k-m-1$ terms in the first line in the right-hand side of eq. (\ref{sk2}) are equal to the left-hand side (without the minus sign), and all the $m$ terms in the second line are equal to each other. Thus, keeping in mind the symmetrization,
\begin{equation}\label{sk3}
  (a_1\cdots a_m [b,c_{m+1}] c_{m+2} \cdots c_k) = - \frac{m}{k-m}(a_1 \cdots [b,a_m] c_{m+1} c_{m+2} \cdots c_k)\;, 
\end{equation}
which proves the identity $({\cal C}^k_m)_{A1} = -({\cal C}^k_m)_{A2}$. 

Next we prove the identity  $({\cal C}^k_m)_{A2} = ({\cal C}^k_m)_{A4} - 2({\cal C}^k_m)_{A5}$. To this end we rewrite the factors $({\cal C}^k_m)_{A5}$ and  $({\cal C}^k_m)_{A2}$ as follows:
\begin{eqnarray}
({\cal C}^k_m)_{A5} &=& -\frac{(a_1\cdots a_{m-2} [b,a_{m-1}] [b,a_{m}] a_{m+1}  \cdots a_k) (c_1\cdots c_k)}{[(k-m)!]^2m!(m-2)!2!} \label{1311}\\
 &\times& (a_1\cdots a_m c_{m+1} \cdots c_k)(c_1\cdots c_m  a_{m+1} \cdots a_k)   \nonumber\\
 ({\cal C}^k_m)_{A2} &=& - \frac{(a_1\cdots a_{m-1} [b,a_m] [b,a_{m+1}] a_{m+2}  \cdots a_k) (c_1\cdots c_k)}{{(k-m)!(k-m-1)!m!(m-1)!}} \nonumber\\
  &\times& (a_1\cdots a_m c_{m+1} \cdots c_k)(c_1\cdots c_m  a_{m+1} \cdots a_k) \label{1312}  
\end{eqnarray}  
Repeating the steps described above and using the identity $[t_b,[t_b,t_{a_m}]] = f_{{a_m}de}f_{deb} t_b$ we obtain
\begin{eqnarray} 
 &&\hskip-1cm -(a_1\cdots a_{m-2} [b,a_{m-1}] [b,a_{m}] a_{m+1}  \cdots a_k) = \nonumber\\
 && {}\frac{f_{{a_m}de}f_{deb}}{m-1} (a_1\cdots a_{m-1} b a_{m+1} \cdots a_k) +  \frac{k-m}{m-1} (a_1\cdots a_{m-1} [b,a_m] [b,a_{m+1}] a_{m+2} \cdots a_k) \nonumber\;. 
\end{eqnarray}
Inserting this into  (\ref{1311}) and recalling (\ref{13}), we obtain the desired result. 

We remark that similar identities among the colour factors should also hold if we replace single-trace by multi-trace operators. In fact, such identities are corollaries of harmonic analyticity, which in turn is a consequence of the BPS shortness conditions.

Finally, we can compute, for instance, the colour factor $({\cal C}^k_m)_{A1}$ in the large $N$ limit. The basic rule are
\begin{eqnarray}
   (a_1 \cdots a_l A) (a_1 \cdots a_l B) &\approx& (l+1)^2 l! N^{l-1} {\rm Tr} AB \nonumber\\
  {\rm Tr}[a_l \cdots a_1 a_1 \cdots a_l A]  &\approx& N^l {\rm Tr}A \;. \nonumber \end{eqnarray}
This easily leads to (\ref{colorn}).

\section{OPE and propagator bases}\label{A}

Here we construct a linear transformation from the OPE basis to the propagator basis.
We assume that the irrep $[a_1,a_2,a_3]$ is described by a $C$-tensor
$C_{i_1...i_m}^I$, where $I$ runs over a basis of the irrep and the fundamental 
SO(6) indices $i_1,...,i_m$ (the number $m$ depends on the representation chosen)
are symmetrized according to the corresponding Young pattern.
The linear transformation from the OPE to the propagator basis can be constructed 
owing to the fact that the sum 
\bea
\label{CC}
C^I_{i_1...i_m}C^I_{j_1...j_m}
\eea
expresses a completeness condition, i.e. it is the identity operator acting in the corresponding representation space. Thus, we can write it as product of Kronecker deltas and subject the indices to the required Young symmetry; the overall normalization is fixed by requiring $C^I_{i_1...i_m}C^J_{i_1...i_m}=\d^{IJ}$. 
Now, substituting the completeness condition (\ref{CC}) into eq. (\ref{OPEbasis})
and performing the contractions of the fundamental indices, we arrive at an expression for eq. (\ref{OPEbasis}) in terms of the elements of the propagator basis. When the dimension of the irrep exchanged is sufficiently large, the explicit expression of the identity operator becomes very involved. One can see however that increasing the external weight $k$ in eq. (\ref{OPEbasis}) by one unit makes {\it only three  new representations} exchanged appear, $[0,2k+2,0]$, $[1,2k,1]$ and $[2,2k-2,2]$.\footnote{Only the irreps $[a_1,a_2,a_3]$ with $a_1=a_3=0,1,2$ participate in the supergravity effective action. The other irreps
entering the tensor product decomposition (\ref{tp}) arise in the $4$-point amplitude derived from the effective action as a result of permuting the external points $x_1$...$x_4$.} The irreps arising in the intermediate channel, which are the same for weights $k$ and $k+1$, will have expansions over the corresponding propagator bases but with the same relative coefficients. Thus, once we know the transformation of the OPE basis to the propagator bases for some external weight $k$, we know it as well for the weight $k+1$ and for all those irreps which are exchanged in the case of weight $k$. The expansions over the propagator structures for the three new irreps indicated above can then be constructed by using the powerful identities (B.5), (B.8)-(B.11) from Ref. \cite{AF2} (see \cite{ADOS} for an example of their application).

Below we summarize our findings for operators of weight 4:

\bea
\nonumber
&&\langle  C^1C^2C^5_{[0,0,0]}\rangle \langle  C^3C^4 C^5_{[0,0,0]}\rangle=\d^{12}\d^{34};
\\
\nonumber
&&\langle  C^1C^2C^5_{[0,2,0]}\rangle \langle  C^3C^4 C^5_{[0,2,0]}\rangle=
\frac{1}{2}C^{1234}+\frac{1}{2}C^{1243}-\frac{1}{6}\d^{12}\d^{34} \, ;   \\
 \nonumber
&&\langle  C^1C^2C^5_{[0,4,0]}\rangle \langle  C^3C^4 C^5_{[0,4,0]}\rangle=
\frac{1}{6}\left(4\Upsilon^{1234}+\Omega^{1234}+\Omega^{1243}\right)  
-\frac{2}{15}\left(C^{1234}+C^{1243}\right)\\
\nonumber
&&\hspace{2.5cm}
+\frac{1}{60}\d^{12}\d^{34} \, ; \\
\nonumber
&&\langle  C^1C^2C^5_{[0,6,0]}\rangle \langle  C^3C^4 C^5_{[0,6,0]}\rangle=
\frac{1}{20}\left( C^{1423}+C^{1324}+9\Upsilon^{1423}+9\Upsilon^{1324}\right) \\
\nonumber
&&\hspace{1.5cm}
-\frac{9}{140}\left(4\Upsilon^{1234}+\Omega^{1234}+\Omega^{1243}\right)
+\frac{3}{140}\left( C^{1234}+C^{1243}\right)
-\frac{1}{700}\d^{12}\d^{34} \, ;  \\
\nonumber
&&\langle  C^1C^2C^5_{[0,8,0]}\rangle \langle  C^3C^4 C^5_{[0,8,0]}\rangle=
-\frac{2}{735}(C^{1234}+C^{1243})-\frac{8}{315}(C^{1324}+C^{1423})\\
\nonumber
&&\hspace{1.5cm}+\frac{8}{35}(C^{1342}+C^{1432})
+\frac{2}{35}\left(\Upsilon^{1234}-4\Upsilon^{1324}-4\Upsilon^{1423} \right)
   \\
\nonumber
&&\hspace{1.5cm}+\frac{1}{70}\left(\Omega^{1234} +\Omega^{1243} +36\Omega^{1342} \right)
+\frac{1}{8820}\d^{12}\d^{34}+\frac{1}{70}(\d^{14}\d^{23}+\d^{13}\d^{24}) \, ;\\
\nonumber
&&\langle  C^1C^2C^5_{[1,0,1]}\rangle \langle  C^3C^4 C^5_{[1,0,1]}\rangle =
2(C^{1234}-C^{1243}) \, ;   \nonumber  
\eea
\bea
&&\langle  C^1C^2C^5_{[1,2,1]}\rangle \langle  C^3C^4 C^5_{[1,2,1]}\rangle =
\frac{1}{3}(-C^{1234}+C^{1243}+2\Omega^{1234}-2\Omega^{1243}) \, ;\nonumber  \\
\nonumber
&&\langle  C^1C^2C^5_{[1,4,1]}\rangle \langle  C^3C^4 C^5_{[1,4,1]}\rangle =
\frac{1}{25}(C^{1234}-C^{1243}-5\Omega^{1234}+5\Omega^{1243}\\
\nonumber
&&\hspace{1.5cm}+15\Upsilon^{1324}-15\Upsilon^{1423}+5C^{1324}-5C^{1423}) \, ;\\
\nonumber
&&\langle  C^1C^2C^5_{[1,6,1]}\rangle \langle  C^3C^4 C^5_{[1,6,1]}\rangle =
\frac{1}{245}\Big(-C^{1234}+C^{1243}+21(C^{1423}-C^{1324}) \\
\nonumber
&&\hspace{0.5cm}+112(C^{1342}-C^{1432}) 
+63(\Upsilon^{1423}-\Upsilon^{1324})+9(\Omega^{1234}-\Omega^{1243})
+14(\d^{13}\d^{24}-\d^{14}\d^{23}) \, 
\Big) \, ; \\
\nonumber
&&\langle  C^1C^2C^5_{[2,0,2]}\rangle \langle  C^3C^4 C^5_{[2,0,2]}\rangle =
-\frac{2}{3}\Big(
C^{1234}+C^{1243}\\
\nonumber
&&\hspace{2.5cm}
-2(\Omega^{1234}+\Omega^{1243})+4\Upsilon^{1234}
-\frac{1}{5}\d^{12}\d^{34} 
\Big) \, ;\\
\nonumber
&&\langle  C^1C^2C^5_{[2,2,2]}\rangle \langle  C^3C^4 C^5_{[2,2,2]}\rangle =
\frac{8}{15}\Big(\frac{2}{7}(C^{1234}+C^{1243})+C^{1324}+C^{1423}\\
\nonumber
&&\hspace{2.5cm}-\Upsilon^{1324}-\Upsilon^{1423}-
\Omega^{1234}-\Omega^{1243}-\frac{1}{35}\d^{12}\d^{34}
\Big) \,  ; \\
\nonumber
&&\langle  C^1C^2C^5_{[2,4,2]}\rangle \langle  C^3C^4 C^5_{[2,4,2]}\rangle =
\frac{1}{735}\Big(
 -17(C^{1234}+C^{1243})-189(C^{1324}+C^{1423})\\
\nonumber
&&\hspace{2.5cm}+252(C^{1432}+C^{1342})+112\Upsilon^{1234}-189(\Upsilon^{1324}+\Upsilon^{1423})
-756\Omega^{1342}
\\
\nonumber
&&\hspace{2.5cm}+91(\Omega^{1234}+\Omega^{1243})+126(\d^{13}\d^{24}+\d^{14}\d^{23})+\d^{12}\d^{34}
\Big) .
\eea
  
To work out the operator product expansion of the four-point amplitude (\ref{Ampl}),
one has to construct the projectors on the different SU(4) channels arising 
in the decomposition (\ref{tp}) (see Refs. \cite{AFP,ADOS}). To this end it is convenient to find out the pairings among the elements of the propagator basis. This can be done by using once again the completeness condition and we summarize the results in the two tables below.

\vskip 1cm
\begin{center}
\begin{tabular}{c|cccccc}
Tensor      & $C^{1234}$ & $C^{1243}$ & $C^{1324}$ & $C^{1342}$ & $C^{1423}$ & $C^{1432}$ \\
\hline
&&&&&&\\
$C^{1234}$  & $\frac{30429}{20}$ & $\frac{1029}{20}$ & $\frac{39039}{100}$& $\frac{1239}{100}$ & 
$\frac{1239}{100}$ &$\frac{189}{100}$ \\
&&&&&&\\
$\Omega^{1234}$  & $\frac{149499}{200}$ & $\frac{2499}{200}$ & $\frac{149499}{200}$& $\frac{2499}{200}$ & 
$\frac{567}{200}$ &$\frac{567}{200}$ \\
&&&&&&\\
$\Upsilon^{1234}$  & $\frac{15729}{200}$ & $\frac{15729}{200}$ & $\frac{6069}{200}$& $\frac{1869}{200}$ & 
$\frac{6969}{200}$ &$\frac{1869}{200}$ \\
&&&&&&\\
$\d^{12}\d^{34}$  & $\frac{3675}{2}$ & $\frac{3675}{2}$ & $\frac{441}{2}$& $\frac{21}{2}$ & 
$\frac{441}{2}$ &$\frac{21}{2}$ \\
&&&&&&\\
$\d^{13}\d^{24}$  & $\frac{441}{2}$ & $\frac{21}{2}$ & $\frac{3675}{2}$& $\frac{3675}{2}$ & 
$\frac{21}{2}$ &$\frac{441}{2}$ \\
&&&&&&\\
$\d^{14}\d^{23}$  & $\frac{21}{2}$ & $\frac{441}{2}$ & $\frac{21}{2}$& $\frac{441}{2}$ & 
$\frac{3675}{2}$ &$\frac{3675}{2}$ 
\end{tabular}
\end{center}

\vskip 1cm
\begin{center}
\begin{tabular}{c|cccccc}
Tensor      & $\Omega^{1234}$ & $\Omega^{1243}$ & $\Omega^{1432}$ & $\Upsilon^{1234}$ & $\Upsilon^{1324}$ & $\Upsilon^{1432}$   \\
\hline
&&&&&&\\
$\Omega^{1234}$  & $\frac{370041}{400}$ & $\frac{777}{400}$ & $\frac{777}{400}$& $\frac{1407}{80}$ & 
$\frac{1407}{80}$ &$\frac{567}{80}$ \\
&&&&&&\\
$\Upsilon^{1234}$  & $\frac{1407}{80}$ & $\frac{1407}{80}$ & $\frac{567}{80}$& $\frac{51681}{400}$ & 
$\frac{13041}{400}$ &$\frac{13041}{400}$ \\
&&&&&&\\
$\d^{12}\d^{34}$  & $\frac{2205}{4}$ & $\frac{2205}{4}$ & $\frac{21}{4}$   & $\frac{735}{4}$ & $\frac{147}{4}$ & $\frac{147}{4}$       \\
&&&&&&\\
$\d^{13}\d^{24}$  & $\frac{2205}{4}$ & $\frac{21}{4}$ & $\frac{2205}{4}$  & $\frac{147}{4}$ & $\frac{735}{4}$ & $\frac{147}{4}$             \\
&&&&&&\\
$\d^{14}\d^{23}$  & $\frac{21}{4}$ & $\frac{2205}{4}$ & $\frac{2205}{4}$   & $\frac{147}{4}$ & $\frac{147}{4}$ & $\frac{735}{4}$    
\end{tabular}
\end{center}

\begin{center}
\begin{tabular}{l}
Tables 2 and 3: The pairings of the various propagator structures. 
The number \\ appearing at the intersection of a row and 
a column is the value of the pairing \\
of the corresponding tensors, e.g., the value 
of $\Omega^{1234}\Upsilon^{1432}$ is 567/80.
\end{tabular}
\end{center}
\vskip 0.5cm

It is worthwhile noting that every number occurring in these tables is a multiple of three. The Tables of pairings can be used firstly to check the orthogonality of the elements of the OPE basis found before, and secondly to establish the projectors on three missing irreps $[4,0,4]$, $[3,0,3]$ and $[3,2,3]$.

The projectors on the irreps $[4,0,4]$,  $[3,0,3]$ and $[3,2,3]$ 
can be found by requiring them to be mutually orthogonal and to be orthogonal to any 
of the twelve other projectors found above.  In this way we obtain
\bea
\nonumber
&&P^{1234}_{[4,0,4]}\sim
14(C^{1234}+C^{1243})-189(\Upsilon^{1324}+\Upsilon^{1423}-C^{1324}-C^{1423})\\
\nonumber
&&\hspace{2.5cm}+504(C^{1432}+C^{1342})+56\Upsilon^{1234}
-756\Omega^{1342}
\\
\nonumber
&&\hspace{2.5cm}-91(\Omega^{1234}+\Omega^{1243})-126(\d^{13}\d^{24}+\d^{14}\d^{23})-\d^{12}\d^{34}\, ; \\
\nonumber
&&P^{1234}_{[3,0,3]}\sim
2(C^{1234}-C^{1243})+7(C^{1324}-C^{1423})-21(\Upsilon^{1324}-\Upsilon^{1423})\\
\nonumber
&&\hspace{2.5cm}
-7(\Omega^{1234}-\Omega^{1243})\,  ; \\
\nonumber
&&P^{1234}_{[3,2,3]}\sim
2(C^{1234}-C^{1243})+27(C^{1324}-C^{1423})+36(C^{1342}-C^{1423})
\\
\nonumber
&&\hspace{2.5cm}-9(\Upsilon^{1324}-\Upsilon^{1423})
-13(\Omega^{1234}-\Omega^{1243})-18(\d^{13}\d^{24}-\d^{14}\d^{23}) \, .
\eea

\section{$D$-functions.}\label{B}
Here we present the coefficient functions of the four-point amplitude for the $\half$-BPS
operators of weight 4 and outline the basic steps of how they can be further 
simplified. 

The original four-point amplitude obtained by summing up different AdS graphs is written in terms 
of the so-called $D$-functions.
The $D$-functions related to $AdS_{d+1}$ are defined by the formula \cite{DHFMMR}
\begin{eqnarray} 
\label{defD} \hskip -0.2cm
D_{\D_1\D_2\D_3\D_4}(x_1,x_2,x_3,x_4)= \int \frac{\rmd^{d} 
w~ \rmd w_0}{w_0^{d+1}} \prod_{i=1}^4 K_{\D_i}(x) \, , \quad K_{\D}(x)
= \left(\frac{w_0}{w_0^2+(\w-x)^2}\right)^{\D} \!  ,
\end{eqnarray} 
where the integral is taken over the space parametrized by $w=(w_0,\w)$, 
$\w$ being a $d$-dimensional vector.
In what follows it is useful to introduce 
the $\oD$-functions \cite{DO} which depend on the conformal cross-ratios $s$ and $t$
and are related to the corresponding $D$-functions as follows:
\bea
\nonumber
 \bar{D}_{\D_1\D_2\D_3\D_4}(s,t)=\frac{2\prod_i\Gamma(\D_i) }{\pi^{d/2}\G(\Sigma-d/2)}
\frac{(x_{13}^2)^{\Sigma-\D_4}(x_{24}^2)^{\D_2}}{(x_{14}^2)^{\Sigma-\D_1-\D_4}(x_{34}^2)^{\Sigma-\D_3-\D_4} }
 D_{\D_1\D_2\D_3\D_4}(x_1,x_2,x_3,x_4),
\eea
where $\Sigma=1/2\sum_i\D_i$. For $d=4$ and $\D_i=1$ we define $\oD_{1111}=\Phi(s,t)$, where $\Phi(s,t)$ is the standard (one-loop) box integral in four dimensions.

In terms of the $\oD$-functions the coefficient functions of the four-point amplitude for the $\half$-BPS operators of weight 4 read as follows:
\bea
\nonumber
a_1&=&-\frac{2}{N^2}s^2\Big[4\oD_{2211}+4\oD_{2222}+4(s-t-1)\oD_{3322}
+2(s-2t-2)\oD_{3333}\\
\nonumber
&+&s(s-4t-4)\oD_{4433}-s(2+5s+2t)\oD_{4444}+s^2(s-t-1)\oD_{5544}
\Big] \, ;\\
\nonumber
b_1&=&\frac{4}{N^2}s^3\Big[9\oD_{4343}+2\oD_{4411}-10\oD_{4422}-9\oD_{4433}+3\oD_{4435}+
12\oD_{4444}+2t\oD_{4521}\\
\nonumber
&+&(1-3s)\oD_{4534}-2t\oD_{4554}-2\oD_{5421}-\oD_{5434}+2(1-s+t)
(\oD_{5533}+\oD_{5544})  
\Big]\, ;
\eea
\bea
\nonumber
c_1&=&\frac{2}{N^2}\Big[16(\oD_{3245}-\oD_{3254}+s^2\oD_{3425}-s^3\oD_{3524})
+17s^2(\oD_{4343}+\oD_{4433})\\
\nonumber
&+&14s(\oD_{4345}-\oD_{4354}+s\oD_{4435}-s^2\oD_{4534})
+32s^2(\oD_{4422}+\oD_{4242})\\
\nonumber
&-&78s^2\oD_{4444}
+s^2(s+t+15)\oD_{5454}+s^2(1+15s+t)\oD_{5544}
\Big] \, ;
\eea
\bea
\nonumber
d_1=\frac{4}{N^2}s^2\Big[81(\oD_{4334}+t\oD_{4343})
+9(t\oD_{4435}+t\oD_{4453}-st\oD_{4534}-st\oD_{4543}) \\
\nonumber
-6t(\oD_{5454}+t\oD_{4554})+64t\oD_{4422}+30t\oD_{4433}+2t(1-s+t)\oD_{5544} 
\Big] \, .
\eea
These expressions are rather involved and to simplify them one has to use 
the various identities between $\oD$'s. We refer the reader to Appendices 
D of Refs. \cite{DO,ADOS} where a complete list of the necessary identities is given 
and the basic technique of their usage is explained 
(see also the Appendix to Ref. \cite{DHFMMR}).  

We start by showing how the coefficient $a_1$ can be simplified.  To this end we 
need to successively use the following formulae:
\bea
\nonumber
(1-s+t)\oD_{5544}&=&-2\oD_{5535}+9\oD_{4433}-7\oD_{4444}\,  \\
\nonumber
(1-s+t)\oD_{4444}&=&-2\oD_{4435}+9\oD_{3333}-6s\oD_{4433} \, \\
\nonumber
s(1-s+t)\oD_{4433}&=&-2\oD_{3335}-5s\oD_{3333}+4\oD_{2233} \, \\
\nonumber
(1-s+t)\oD_{3333}&=&-2\oD_{3324}-4\oD_{2233}+4\oD_{2222}\, \\
\nonumber
(1-s+t)\oD_{3322}&=&-2\oD_{3313}+\oD_{2211}-3\oD_{2222} \, .
\eea 
After this we can make the further substitution 
\bea
\nonumber
s\oD_{4435}&=&\oD_{3346}-\oD_{3335}\,  \\
\nonumber
s\oD_{3324}&=&\oD_{2235}-\oD_{2224}\,  \\
\nonumber
s\oD_{3335}&=&\oD_{2246}-2\oD_{2235}  
\eea
and $s\oD_{3313}=\oD_{2224}$, $s\oD_{5535}=\oD_{4446}$. In this way we obtain the formula 
\bea
a_1=-\frac{4}{N^2}\Big(2s\oD_{2246}+2s^2\oD_{3346}+s^3\oD_{4446}\Big) \, 
\eea
from which we read off the function $\cF$ (\ref{simpleFG}).

Now we briefly describe the procedure for obtaining the second function $\cG$
from the coefficient $b_1$. This time we need the following identities:
\bea
\nonumber
(t-s-1)\oD_{6422}&=&2(\oD_{5423}-\oD_{5412}-2\oD_{5322})+\frac{2}{s^3}\\
\nonumber
(t-s-1)\oD_{6433}&=&\oD_{5414}-3\oD_{5423}+2\oD_{5434}-4\oD_{5333}\\
\nonumber
(t-s-1)\oD_{6444}&=&2\oD_{5445}-4(\oD_{5434}+\oD_{5344})+2\oD_{5425}
\eea
which imply 
\bea
\label{fp}
&&(t-s-1)\cF=\\
\nonumber
&&-\frac{8}{N^2}s^3\left(\oD_{5445}-\oD_{5423}+\oD_{5425}+\oD_{5414}-2\oD_{5344}
-2\oD_{5412}-4\oD_{5333}-4\oD_{5322}\right)-\frac{16}{N^2}
\eea
Substituting this into 
\bea
\nonumber
\cG \equiv \beta_3=\frac{1}{s}(b_1-(t-s-1)\cF)
\eea
and making again use of the identities (D.7)-(D.11) from Ref. \cite{ADOS} we arrive at our final expression (\ref{simpleFG}).

\renewcommand{\baselinestretch}{0.6}

\end{document}

%% file: fig4.pstex_t
\begin{picture}(0,0)%
\special{psfile=fig4.pstex}%
\end{picture}%
\setlength{\unitlength}{1579sp}%
\begingroup\makeatletter\ifx\SetFigFont\undefined
\def\x#1#2#3#4#5#6#7\relax{\def\x{#1#2#3#4#5#6}}%
\expandafter\x\fmtname xxxxxx\relax \def\y{splain}%
\ifx\x\y   
\gdef\SetFigFont#1#2#3{%
  \ifnum #1<17\tiny\else \ifnum #1<20\small\else
  \ifnum #1<24\normalsize\else \ifnum #1<29\large\else
  \ifnum #1<34\Large\else \ifnum #1<41\LARGE\else
     \huge\fi\fi\fi\fi\fi\fi
  \csname #3\endcsname}%
\else
\gdef\SetFigFont#1#2#3{\begingroup
  \count@#1\relax \ifnum 25<\count@\count@25\fi
  \def\x{\endgroup\@setsize\SetFigFont{#2pt}}%
  \expandafter\x
    \csname \romannumeral\the\count@ pt\expandafter\endcsname
    \csname @\romannumeral\the\count@ pt\endcsname
  \csname #3\endcsname}%
\fi
\fi\endgroup
\begin{picture}(10990,16893)(226,-16120)
\end{picture}

%% file: frule1.pstex_t
\begin{picture}(0,0)%
\special{psfile=frule1.pstex}%
\end{picture}%
\setlength{\unitlength}{3947sp}%
\begingroup\makeatletter\ifx\SetFigFont\undefined
\def\x#1#2#3#4#5#6#7\relax{\def\x{#1#2#3#4#5#6}}%
\expandafter\x\fmtname xxxxxx\relax \def\y{splain}%
\ifx\x\y   
\gdef\SetFigFont#1#2#3{%
  \ifnum #1<17\tiny\else \ifnum #1<20\small\else
  \ifnum #1<24\normalsize\else \ifnum #1<29\large\else
  \ifnum #1<34\Large\else \ifnum #1<41\LARGE\else
     \huge\fi\fi\fi\fi\fi\fi
  \csname #3\endcsname}%
\else
\gdef\SetFigFont#1#2#3{\begingroup
  \count@#1\relax \ifnum 25<\count@\count@25\fi
  \def\x{\endgroup\@setsize\SetFigFont{#2pt}}%
  \expandafter\x
    \csname \romannumeral\the\count@ pt\expandafter\endcsname
    \csname @\romannumeral\the\count@ pt\endcsname
  \csname #3\endcsname}%
\fi
\fi\endgroup
\begin{picture}(2460,409)(-2789,7668)
\put(-329,7739){\makebox(0,0)[lb]{\smash{\SetFigFont{12}{14.4}{rm}$\langle \tilde q^+_a(1) q^+_b(2)\rangle|_{\theta=0} = -\frac{\delta_{ab}}{4\pi^2} \frac{[12]}{x^2_{12}} $}}}
\put(-2789,7889){\makebox(0,0)[lb]{\smash{\SetFigFont{12}{14.4}{rm}$1a$}}}
\put(-1544,7889){\makebox(0,0)[lb]{\smash{\SetFigFont{12}{14.4}{rm}$2b$}}}
\end{picture}

%% file: frule2.pstex_t
\begin{picture}(0,0)%
\special{psfile=frule2.pstex}%
\end{picture}%
\setlength{\unitlength}{2368sp}%
\begingroup\makeatletter\ifx\SetFigFont\undefined
\def\x#1#2#3#4#5#6#7\relax{\def\x{#1#2#3#4#5#6}}%
\expandafter\x\fmtname xxxxxx\relax \def\y{splain}%
\ifx\x\y   
\gdef\SetFigFont#1#2#3{%
  \ifnum #1<17\tiny\else \ifnum #1<20\small\else
  \ifnum #1<24\normalsize\else \ifnum #1<29\large\else
  \ifnum #1<34\Large\else \ifnum #1<41\LARGE\else
     \huge\fi\fi\fi\fi\fi\fi
  \csname #3\endcsname}%
\else
\gdef\SetFigFont#1#2#3{\begingroup
  \count@#1\relax \ifnum 25<\count@\count@25\fi
  \def\x{\endgroup\@setsize\SetFigFont{#2pt}}%
  \expandafter\x
    \csname \romannumeral\the\count@ pt\expandafter\endcsname
    \csname @\romannumeral\the\count@ pt\endcsname
  \csname #3\endcsname}%
\fi
\fi\endgroup
\begin{picture}(9024,2738)(2389,1439)
\put(9211,2249){\makebox(0,0)[lb]{\smash{\SetFigFont{7}{8.4}{rm}${\rm Tr}(W^2)$}}}
\put(11101,3539){\makebox(0,0)[lb]{\smash{\SetFigFont{7}{8.4}{rm}$q$}}}
\put(7801,3539){\makebox(0,0)[lb]{\smash{\SetFigFont{7}{8.4}{rm}$\tilde q$}}}
\put(3601,2249){\makebox(0,0)[lb]{\smash{\SetFigFont{7}{8.4}{rm}$W$}}}
\put(5701,3539){\makebox(0,0)[lb]{\smash{\SetFigFont{7}{8.4}{rm}$q$}}}
\put(2401,3539){\makebox(0,0)[lb]{\smash{\SetFigFont{7}{8.4}{rm}$\tilde q$}}}
\put(11026,3989){\makebox(0,0)[lb]{\smash{\SetFigFont{7}{8.4}{rm}$2c$}}}
\put(7801,3989){\makebox(0,0)[lb]{\smash{\SetFigFont{7}{8.4}{rm}$1a$}}}
\put(9526,2864){\makebox(0,0)[lb]{\smash{\SetFigFont{7}{8.4}{rm}$0$}}}
\put(2401,3989){\makebox(0,0)[lb]{\smash{\SetFigFont{7}{8.4}{rm}$1a$}}}
\put(5626,3989){\makebox(0,0)[lb]{\smash{\SetFigFont{7}{8.4}{rm}$2c$}}}
\put(4426,2264){\makebox(0,0)[lb]{\smash{\SetFigFont{7}{8.4}{rm}$0b$}}}
\end{picture}

%% file: diag2.pstex_t
\begin{picture}(0,0)%
\special{psfile=diag2.pstex}%
\end{picture}%
\setlength{\unitlength}{1973sp}%
\begingroup\makeatletter\ifx\SetFigFont\undefined
\def\x#1#2#3#4#5#6#7\relax{\def\x{#1#2#3#4#5#6}}%
\expandafter\x\fmtname xxxxxx\relax \def\y{splain}%
\ifx\x\y   
\gdef\SetFigFont#1#2#3{%
  \ifnum #1<17\tiny\else \ifnum #1<20\small\else
  \ifnum #1<24\normalsize\else \ifnum #1<29\large\else
  \ifnum #1<34\Large\else \ifnum #1<41\LARGE\else
     \huge\fi\fi\fi\fi\fi\fi
  \csname #3\endcsname}%
\else
\gdef\SetFigFont#1#2#3{\begingroup
  \count@#1\relax \ifnum 25<\count@\count@25\fi
  \def\x{\endgroup\@setsize\SetFigFont{#2pt}}%
  \expandafter\x
    \csname \romannumeral\the\count@ pt\expandafter\endcsname
    \csname @\romannumeral\the\count@ pt\endcsname
  \csname #3\endcsname}%
\fi
\fi\endgroup
\begin{picture}(9450,16389)(1201,-7711)
\put(7351,5789){\makebox(0,0)[lb]{\smash{\SetFigFont{6}{7.2}{rm}$e$}}}
\put(2401,8489){\makebox(0,0)[lb]{\smash{\SetFigFont{6}{7.2}{rm}$a_1$}}}
\put(5251,6389){\makebox(0,0)[lb]{\smash{\SetFigFont{6}{7.2}{rm}$c_k$}}}
\put(9301,4589){\makebox(0,0)[lb]{\smash{\SetFigFont{6}{7.2}{rm}$c_1$}}}
\put(9301,5189){\makebox(0,0)[lb]{\smash{\SetFigFont{6}{7.2}{rm}$c_m$}}}
\put(9601,5639){\makebox(0,0)[lb]{\smash{\SetFigFont{6}{7.2}{rm}$c_{m+1}$}}}
\put(6601,7289){\makebox(0,0)[lb]{\smash{\SetFigFont{6}{7.2}{rm}$a_k$}}}
\put(7351,7589){\makebox(0,0)[lb]{\smash{\SetFigFont{6}{7.2}{rm}$a_{m+1}$}}}
\put(7801,7889){\makebox(0,0)[lb]{\smash{\SetFigFont{6}{7.2}{rm}$a_m$}}}
\put(4351,2339){\makebox(0,0)[lb]{\smash{\SetFigFont{6}{7.2}{rm}$d$}}}
\put(3901,-811){\makebox(0,0)[lb]{\smash{\SetFigFont{6}{7.2}{rm}$c_1$}}}
\put(3901,-211){\makebox(0,0)[lb]{\smash{\SetFigFont{6}{7.2}{rm}$c_m$}}}
\put(4201, 89){\makebox(0,0)[lb]{\smash{\SetFigFont{6}{7.2}{rm}$c_{m+1}$}}}
\put(4501,1589){\makebox(0,0)[lb]{\smash{\SetFigFont{6}{7.2}{rm}$e$}}}
\put(1201,1889){\makebox(0,0)[lb]{\smash{\SetFigFont{6}{7.2}{rm}$a_k$}}}
\put(1951,2189){\makebox(0,0)[lb]{\smash{\SetFigFont{6}{7.2}{rm}$a_{m+1}$}}}
\put(2251,2489){\makebox(0,0)[lb]{\smash{\SetFigFont{6}{7.2}{rm}$a_m$}}}
\put(7351,2399){\makebox(0,0)[lb]{\smash{\SetFigFont{6}{7.2}{rm}$a_m$}}}
\put(7321,1889){\makebox(0,0)[lb]{\smash{\SetFigFont{6}{7.2}{rm}$a_{m+1}$}}}
\put(6541,1889){\makebox(0,0)[lb]{\smash{\SetFigFont{6}{7.2}{rm}$a_k$}}}
\put(9601,239){\makebox(0,0)[lb]{\smash{\SetFigFont{6}{7.2}{rm}$c_{m+1}$}}}
\put(9301,-211){\makebox(0,0)[lb]{\smash{\SetFigFont{6}{7.2}{rm}$c_m$}}}
\put(9301,-811){\makebox(0,0)[lb]{\smash{\SetFigFont{6}{7.2}{rm}$c_1$}}}
\put(7801,8489){\makebox(0,0)[lb]{\smash{\SetFigFont{6}{7.2}{rm}$a_1$}}}
\put(2401,3089){\makebox(0,0)[lb]{\smash{\SetFigFont{6}{7.2}{rm}$a_1$}}}
\put(7801,3089){\makebox(0,0)[lb]{\smash{\SetFigFont{6}{7.2}{rm}$a_1$}}}
\put(3301,6839){\makebox(0,0)[lb]{\smash{\SetFigFont{6}{7.2}{rm}$b$}}}
\put(8311,7019){\makebox(0,0)[lb]{\smash{\SetFigFont{6}{7.2}{rm}$b$}}}
\put(2401,7889){\makebox(0,0)[lb]{\smash{\SetFigFont{6}{7.2}{rm}$a_m$}}}
\put(1951,7589){\makebox(0,0)[lb]{\smash{\SetFigFont{6}{7.2}{rm}$a_{m+1}$}}}
\put(1201,7289){\makebox(0,0)[lb]{\smash{\SetFigFont{6}{7.2}{rm}$a_k$}}}
\put(4501,6989){\makebox(0,0)[lb]{\smash{\SetFigFont{6}{7.2}{rm}$e$}}}
\put(4201,5639){\makebox(0,0)[lb]{\smash{\SetFigFont{6}{7.2}{rm}$c_{m+1}$}}}
\put(3901,5189){\makebox(0,0)[lb]{\smash{\SetFigFont{6}{7.2}{rm}$c_m$}}}
\put(3901,4589){\makebox(0,0)[lb]{\smash{\SetFigFont{6}{7.2}{rm}$c_1$}}}
\put(1951,5789){\makebox(0,0)[lb]{\smash{\SetFigFont{6}{7.2}{rm}$d$}}}
\put(9751,7799){\makebox(0,0)[lb]{\smash{\SetFigFont{6}{7.2}{rm}$d$}}}
\put(10651,6389){\makebox(0,0)[lb]{\smash{\SetFigFont{6}{7.2}{rm}$c_k$}}}
\put(5251,989){\makebox(0,0)[lb]{\smash{\SetFigFont{6}{7.2}{rm}$c_k$}}}
\put(10651,989){\makebox(0,0)[lb]{\smash{\SetFigFont{6}{7.2}{rm}$c_k$}}}
\put(3541,1589){\makebox(0,0)[lb]{\smash{\SetFigFont{6}{7.2}{rm}$b$}}}
\put(5011,-3721){\makebox(0,0)[lb]{\smash{\SetFigFont{6}{7.2}{rm}$a_m$}}}
\put(4681,-4021){\makebox(0,0)[lb]{\smash{\SetFigFont{6}{7.2}{rm}$a_{m+1}$}}}
\put(3931,-4321){\makebox(0,0)[lb]{\smash{\SetFigFont{6}{7.2}{rm}$a_k$}}}
\put(6931,-5971){\makebox(0,0)[lb]{\smash{\SetFigFont{6}{7.2}{rm}$c_{m+1}$}}}
\put(6631,-6421){\makebox(0,0)[lb]{\smash{\SetFigFont{6}{7.2}{rm}$c_m$}}}
\put(6631,-7021){\makebox(0,0)[lb]{\smash{\SetFigFont{6}{7.2}{rm}$c_1$}}}
\put(6781,-3271){\makebox(0,0)[lb]{\smash{\SetFigFont{6}{7.2}{rm}$d$}}}
\put(6631,-3721){\makebox(0,0)[lb]{\smash{\SetFigFont{6}{7.2}{rm}$e$}}}
\put(7981,-5221){\makebox(0,0)[lb]{\smash{\SetFigFont{6}{7.2}{rm}$c_k$}}}
\put(4861,-3301){\makebox(0,0)[lb]{\smash{\SetFigFont{6}{7.2}{rm}$a_{m-1}$}}}
\put(4561,-3001){\makebox(0,0)[lb]{\smash{\SetFigFont{6}{7.2}{rm}$a_1$}}}
\put(9751,2399){\makebox(0,0)[lb]{\smash{\SetFigFont{6}{7.2}{rm}$b$}}}
\put(6151,-3286){\makebox(0,0)[lb]{\smash{\SetFigFont{6}{7.2}{rm}$b$}}}
\end{picture}